\documentclass[review,3p,times]{elsarticle}
\usepackage{amsmath,hyperref}
\usepackage{lineno}
\usepackage{color}
\usepackage{subfigure}
\usepackage{epsfig}
\usepackage{caption,xcolor,float,graphicx}
\usepackage{graphics}
\usepackage{epstopdf}
\biboptions{sort&compress}
\journal{Journal of Colloid and Interface Science}

\begin{document}

\begin{frontmatter}
	
\title{Droplet impact behavior on a hydrophobic plate with a wettability-patterned orifice}
\author[mymainaddress]{Juanyong Wang}

\author[mymainaddress]{Lei Wang\corref{mycorrespondingauthor}}
\cortext[mycorrespondingauthor]{Corresponding author}
\ead{wangleir1989@126.com}
\address[mymainaddress]{School of Mathematics and Physics, China University of Geosciences, Wuhan 430074, China}

\author[mysecondaddress]{Jiangxu Huang}
\address[mysecondaddress]{School of Mathematics and Statistics, Huazhong University of Science and Technology, Wuhan 430074, China}

\author[mythirdaddress]{Dinggen Li}
\address[mythirdaddress]{School of Energy and Power Engineering, Huazhong University of Science and Technology, Wuhan, 430074, China}

\begin{abstract}
Droplet impact behavior has attracted much attention recently due to its academic significance and diverse industrial applications. This study employs the lattice Boltzmann method to simulate the impact of a droplet on a hydrophobic plate featuring a square orifice. Unlike previous studies, the chemical property of the orifice considered in this work is not homogeneous but heterogeneous, and its cross-sectional wettability changes from hydrophobicity to hydrophilicity. The study first validates the numerical method against experimental data, and then investigates in detail the influences of the Weber number, wettability difference, and pore size. According to the numerical results, we observed that the evolutionary stages of the impinging droplet always include the spreading phase and the rebounding phase, while whether there exists the splitting phase, it depends on the combined effect of the wettability difference and the Weber number. The impact behavior of droplets is analyzed by evaluating the underlying mechanisms such as kinetic energy, surface energy, viscous dissipation energy, and pressure. It is interesting to note that the existence of wettability-patterned pore tends to promote adhesion of droplets on the plate, resulting in the droplet impact behaviors are largely different from that for the case of homogeneous pore. Additionally, a phase diagram is constructed for various Weber numbers and pore sizes, revealing that the dynamic behavior of droplets is determined by the competition among dynamic pressure, capillary pressure, and viscous pressure losses. These insights from numerical studies guide the development of innovative solid substrates capable of manipulating droplet motion.

\end{abstract}
	
\end{frontmatter}

\section{Introduction}\label{section1}

The impact of droplets on solid substrates is usually encountered in nature and industry, including inkjet printing \cite{Lohse_ARFM2022}, surface coating \cite{Park_Nature review2020}, and droplet-based microfluidics processes \cite{Moragues_Nature review2023}, et al.. Given its crucial role in numerous practical applications, researchers have extensively studied the behavior of droplet impact under various conditions \cite{Yarin_ARFM2006,Howland_PRL2016}. In the literatures, the most widely used configuration is the droplet impacts on a flat plate \cite{Tran_SoftMatter2013}. However, the flat plate employed in these works is an ideal assumption. In fact, in some industrial applications such as fuel spray \cite{Wang_Scientific Reports2017}, drug production \cite{Norman_2017}, and cell microarrays \cite{Jonczyk_Microarrays2016}, droplet impacting on a perforated plate is commonly seen. Intuitively, the droplet dynamics depend on the surface geometry, and it is expected that the droplet impact behavior in such a case is different from that on a flat plate. Therefore, exploring the impact behavior of droplets on perforated plates holds significant scholarly and pragmatic value.

In recent times, various experimental and numerical endeavors have been devoted to investigating the droplet impact dynamics on a perforated plate. In a study conducted by Lorenceau et al. \cite{Lorenceau2003}, the behavior of droplets impacting on a perforated plate was experimentally analyzed. The study revealed that when the velocity of the droplet surpasses a specific critical value, the droplet will be ejected from the pore. Delbos et al. \cite{Delbos2010} conducted an experimental investigation exploring the forced impregnation of a capillary tube caused by droplet impact and observed the different droplet regimes under various capillary wettability conditions. Bordoloi et al. \cite{Bordoloi2014} experimentally investigated the motion of gravity-driven droplets through a confining orifice, and it is noted that when the orifice surface is hydrophilic, a portion of the droplet becomes trapped as it spreads across both upper and lower surfaces of the plate. Apart from the experimental study, various numerical investigations on droplet impact dynamics on a perforated plate have also been reported. For example, the problem of droplet dripping through a hole is numerically studied by Haghani et al. \cite{Haghani2013}, and the authors found that the gravitational force and adhesion force are two opposing forces in influencing droplet behavior. With a similar configuration, the authors observed four typical deformations of a droplet dripping down a hole, i.e., equilibrium drop on the top of the surface, equilibrium drop under the bottom of the surface, splashing and dripping of the drop \cite{Haghani2016}. Recently, Wang et al. \cite{Wang_PRF_2020} conducted a numerical investigation on the impact of droplets on hydrophobic plates that feature cylindrical pores. They meticulously studied the effects of pore size, plate thickness, and Weber number.

Although the droplet impact behavior on a perforated plate has drawn much attention in the preceding researches, the pore wettability in these works are confined to be homogeneous. Actually, various works have reported that the wettability-patterned surface has a vital effect on the droplet impact behavior. For instance, Kim et al. \cite{Kim_JFM2013} mentally explored the impact of a droplet on a superhydrophobic surface. They compared a surface with a superhydrophilic annulus to a uniformly superhydrophobic surface; authors observed the formation of a water ring followed by the liquid ejecting from a thin lens. Raman et al. \cite{Raman_IJHMT_2016} numerically investigated a droplet impinging on a solid substrate with a wettability difference, they pointed out that the transfer of the momentum from a liquid drop changes as the substrate's wetting properties vary, causing the droplet's momentum to shift from vertical to horizontal. In a study by Huang et al. \cite{Huang_CF2022}, three-dimensional numerical simulations were conducted to examine the impact of double droplets on a wettability-patterned surface. The study suggests that droplets may either rebound or migrate towards regions with higher wettability due to an imbalanced Young's force resulting from wettability difference. In addition, Yang et al. \cite{Yang_JCIS2022} conducted a study on droplet impact on surfaces having mixed wettability. Their study involved both experimental and numerical approaches. They found that the wettability gradient of the surface played a significant role in determining the direction of droplet transport. In a more recent study, Zhang et al. \cite{Zhang_JFM2022} conducted a numerical study on the impact of droplets on a mobile solid surface with heterogeneous properties. They observed that the lateral velocity of the rebound droplet changes by regulating the distance between the impact point and the hydrophilic stripe.

From the above-published literatures, it is evident that the impact dynamics rely on surface geometry and wettability. More recently, porous materials such as textile fabrics and nanofibrous membranes have been reported to have directional fluid transport properties, where the fluid can penetrate through the material from one side to another. Still, its transport is blocked in the reverse direction \cite{Zhou_Scientific reports2013,Zhao_Small_2017}. After a careful revisiting of this material, it is found that directional fluid motion is largely related to the gradient hydrophobic/hydrophilic wettability across the thickness \cite{Zhou_Scientific reports2013,Zhao_Small_2017}. Unfortunately, due to limitations in measurement techniques and existing theoretical models, the droplet impact behavior on this novel material remains unclear.
	
In an endeavor towards bridging current knowledge gaps, the present paper aims to numerically study the droplet impact behavior on a hydrophobic plate with a wettability-patterned orifice. In terms of the numerical method, the lattice Boltzmann (LB) method is adopted, which has witnessed burgeoning growth for simulating complex hydrodynamic phenomena \cite{Aidun_ARFM2010,Kruger_2017,Wang_AMM2019}, as well as the multiphase flows \cite{Li_PECS2016,Chen_IJHMT2014,Liang_AMM2019}. The paper's outline is as follows: The physical problem will be presented in the next section (\ref{section2}), followed by the pseudopotential multiphase LB method in Section \ref{section3}. In section \ref{section4}, we will validate the numerical method against experimental data. Section \ref{section5} presents numerical results and discussion, and a brief conclusion will follow in section \ref{section6}.

\section{Problem statement}\label{section2}

\begin{figure}[H]
	\centering
	\includegraphics[width=1.0\textwidth]{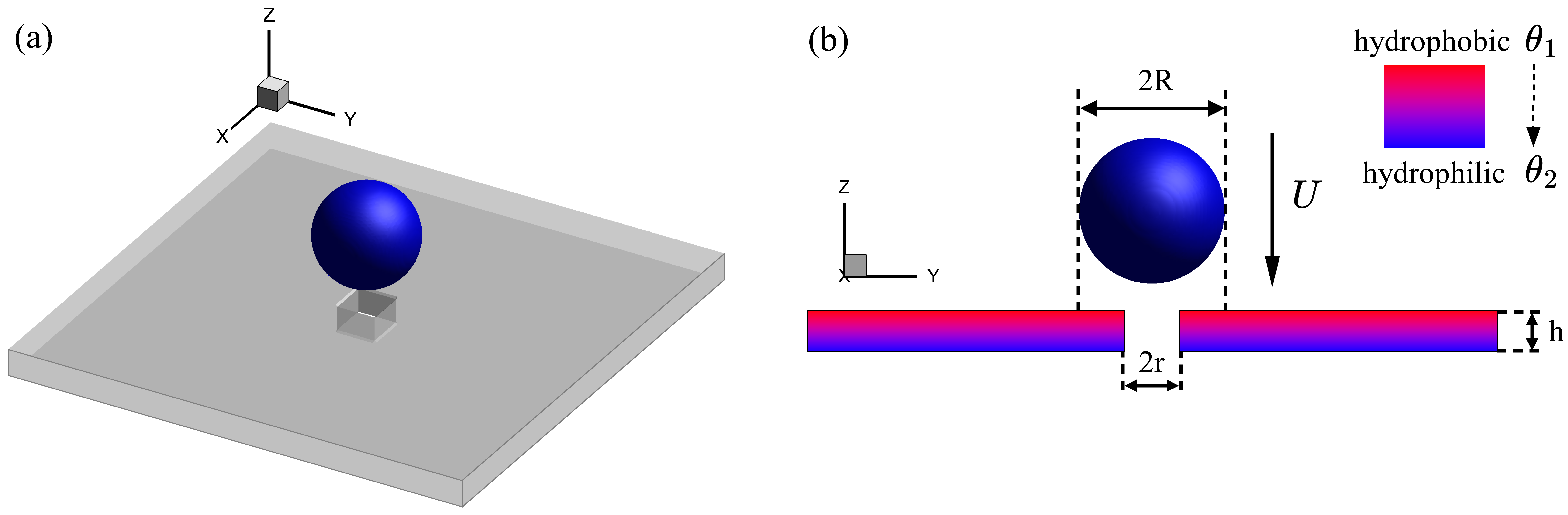}
	\caption{The 3D main view (a) and side view (b) of a droplet impacting on a hydrophobic plate with a wettability-patterned pore.}
	\label{fig1}
\end{figure}

Fig. \ref{fig1} shows a droplet impacting on a hydrophobic plate with a wettability-patterned orifice. The droplet has an initial velocity of $U$ and a radius of $R$ while located above the plate. $h$ represents the plate thickness, and a square with the side length of $2r$ is located in the center of the plate. To investigate the influence of gradient wetting across the thickness, the pore wettability along the cross section changes linearly from hydrophobicity to hydrophilicity, and the contact angles at the top and bottom plates are ${\theta_1}$ and ${\theta_2}$, respectively. With this configuration, the droplet impact dynamics can be described by four dimensionless numbers, i.e., the Weber number ($We$), the Reynolds number ($Re$), the Ohnesorge number ($Oh$), and the capillary number ($Ca$), which are defined as
\begin{equation} \label{eq1}
W e=\frac{2\rho_{L} R U^{2}}{\sigma},  \quad	R e=\frac{2\rho_{L} R U}{\mu_{L}},  \quad	Oh = \frac{{{\mu _L}}}{{\sqrt {2 R \sigma {\rho _L}} }},  \quad	Ca = \frac{{{\mu _L}RU}}{{\sigma r}},
\end{equation} 
in which $\rho_{L}$, $\sigma$, and  $\mu_{L}$ are liquid density, surface tension, and the liquid dynamic viscosity, respectively. Please note that, due to the primary influence of capillary pressure on the impact dynamics in this study, we intentionally set the gravitational acceleration to zero.
 
\section{Pseudopotential multiphase lattice Boltzmann method}
\label{section3}

Several multiphase LB models have been proposed since the introduction of the Lattice Boltzmann (LB) method \cite{Li_PECS2016,Connington_MST2012,Wang_Capillarity2019}. In this work, we adopt the pseudopotential method to model the multiphase flow \cite{Chen_IJHMT2014,Nemati_Physica A2018}, and the distinct advantage of this method is its intrinsic simplicity and mesoscopic nature \cite{Li_PECS2016,Li_CF_2019}. The multiple-relaxation-time (MRT) collision operator is used in the current paper to improve numerical stability. In particular, to facilitate an efficient numerical implementation, a non-orthogonal MRT collision operator is utilized instead of the conventional orthogonal MRT collision operator \cite{Kruger_2017}. Then, the evolution equation in discrete velocity space can be written as \cite{Premnath_JCP2007,Li_CF_2019}
\begin{equation}\label{eq2}
	\begin{aligned} f_{\alpha}\left(\mathbf{x}+\mathbf{e}_{\alpha} \delta_{t}, t+\delta_{t}\right)= & f_{\alpha}(\mathbf{x}, t)-\left.\bar{\Lambda}_{\alpha \beta}\left(f_{\beta}-f_{\beta}^{\mathrm{eq}}\right)\right|_{(\mathbf{x}, t)}+\left.\delta_{t}\left(S_{\alpha}-0.5 \bar{\Lambda}_{\alpha \beta} S_{\beta}\right)\right|_{(\mathbf{x}, t)}\end{aligned},
\end{equation}
in which $f_{\alpha}$ is the density distribution function with velocity at spatial position ${\bf{x}}$ and time $t$, $f_\beta ^{{\rm{eq}}}$ is the equilibrium density distribution function,  ${\delta _t}$ represents the time step, ${S_\beta }$ denotes the forcing term in the velocity space. In addition, $\mathbf{e}_{\alpha}$ is the discrete velocity along the $\alpha$-th direction. In this work, the classical D3Q19 lattice model in three-dimensional space is adopted, and it can be defined as 
\begin{equation}\label{eq3}
	\mathbf{e}_{\alpha}=\left[\begin{array}{lllllllllllllllllll}0 & 1 & -1 & 0 & 0 & 0 & 0 & 1 & -1 & 1 & -1 & 1 & -1 & 1 & -1 & 0 & 0 & 0 & 0 \\ 0 & 0 & 0 & 1 & -1 & 0 & 0 & 1 & -1 & -1 & 1 & 0 & 0 & 0 & 0 & 1 & -1 & 1 & -1 \\ 0 & 0 & 0 & 0 & 0 & 1 & -1 & 0 & 0 & 0 & 0 & 1 & -1 & -1 & 1 & 1 & -1 & -1 & 1\end{array}\right].
\end{equation}
Moreover, $\bar{\Lambda}_{\alpha \beta}=\left(\mathbf{M}^{-1} \mathbf{\Lambda} \mathbf{M}\right)_{\alpha \beta}$ is the collision matrix with ${\bf{M}}$ being the nonorthogonal transformation matrix, and it is given by 
\begin{equation}\label{eq4}
	{\bf{M}} = \left[ {\begin{array}{*{20}{r}}
			1&1&1&1&1&1&1&1&1&1&1&1&1&1&1&1&1&1&1\\
			0&1&{ - 1}&0&0&0&0&1&{ - 1}&1&{ - 1}&1&{ - 1}&1&{ - 1}&0&0&0&0\\
			0&0&0&1&{ - 1}&0&0&1&{ - 1}&{ - 1}&1&0&0&0&0&1&{ - 1}&1&{ - 1}\\
			0&0&0&0&0&1&{ - 1}&0&0&0&0&1&{ - 1}&{ - 1}&1&1&{ - 1}&{ - 1}&1\\
			0&1&1&1&1&1&1&2&2&2&2&2&2&2&2&2&2&2&2\\
			0&2&2&{ - 1}&{ - 1}&{ - 1}&{ - 1}&1&1&1&1&1&1&1&1&{ - 2}&{ - 2}&{ - 2}&{ - 2}\\
			0&0&0&1&1&{ - 1}&{ - 1}&1&1&1&1&{ - 1}&{ - 1}&{ - 1}&{ - 1}&0&0&0&0\\
			0&0&0&0&0&0&0&1&1&{ - 1}&{ - 1}&0&0&0&0&0&0&0&0\\
			0&0&0&0&0&0&0&0&0&0&0&1&1&{ - 1}&{ - 1}&0&0&0&0\\
			0&0&0&0&0&0&0&0&0&0&0&0&0&0&0&1&1&{ - 1}&{ - 1}\\
			0&0&0&0&0&0&0&1&{ - 1}&{ - 1}&1&0&0&0&0&0&0&0&0\\
			0&0&0&0&0&0&0&1&{ - 1}&1&{ - 1}&0&0&0&0&0&0&0&0\\
			0&0&0&0&0&0&0&0&0&0&0&1&{ - 1}&{ - 1}&1&0&0&0&0\\
			0&0&0&0&0&0&0&0&0&0&0&1&{ - 1}&1&{ - 1}&0&0&0&0\\
			0&0&0&0&0&0&0&0&0&0&0&0&0&0&0&1&{ - 1}&{ - 1}&1\\
			0&0&0&0&0&0&0&0&0&0&0&0&0&0&0&1&{ - 1}&1&{ - 1}\\
			0&0&0&0&0&0&0&1&1&1&1&0&0&0&0&0&0&0&0\\
			0&0&0&0&0&0&0&0&0&0&0&1&1&1&1&0&0&0&0\\
			0&0&0&0&0&0&0&0&0&0&0&0&0&0&0&1&1&1&1
	\end{array}} \right].
\end{equation}
Multiplying the transformation matrix ${\bf{M}}$ on the left-hand side of Eq. (\ref{eq2}), the collision process in the moment space can then be expressed as  
\begin{equation}\label{eq5}
	\mathbf{m}^{*}=\mathbf{m}-\boldsymbol{\Lambda}\left(\mathbf{m}-\mathbf{m}^{\mathrm{eq}}\right)+\delta_{t}\left(\mathbf{I}-\frac{\boldsymbol{\Lambda}}{2}\right) \overline{\mathbf{S}},
\end{equation}
in which $\mathbf{m}=\mathbf{M f}$, $\mathbf{m}^{\mathrm{eq}}=\mathbf{M} \mathbf{f}^{\mathrm{eq}}$ are the distribution function and the equilibrium distribution function with $\mathbf{f}$, $\mathbf{f}^{\mathrm{eq}}$ given by ${\bf{f}} = {\left[ {{f_0},{f_1},{f_2},...,{f_{16}},{f_{17}},{f_{18}}} \right]^{\rm T}}$, ${{\bf{f}}^{{\rm{eq}}}} = {\left[ {f_0^{{\rm{eq}}},f_1^{{\rm{eq}}},f_2^{{\rm{eq}}},...,f_{16}^{{\rm{eq}}},f_{17}^{{\rm{eq}}},f_{18}^{{\rm{eq}}}} \right]^{\rm T}}$.
In such a case, $\mathbf{m}^{\mathrm{eq}}$ can be further given as
\begin{equation}\label{eq6}
\begin{array}{*{20}{c}}
	{{{\bf{m}}^{{\rm{eq}}}} = \left[ {\rho ,\rho {u_x},\rho {u_y},\rho {u_z},\rho  + \rho {{\left| {\bf{u}} \right|}^2},\rho \left( {2u_x^2 - u_y^2 - u_z^2} \right),\rho \left( {u_y^2 - u_z^2} \right),\rho {u_x}{u_y},\rho {u_x}{u_z},\rho {u_y}{u_z},\rho c_s^2{u_y},} \right.}\\
	{{{\left. {\rho c_s^2{u_x},\rho c_s^2{u_z},\rho c_s^2{u_x},\rho c_s^2{u_z},\rho c_s^2{u_y},\varphi  + \rho c_s^2\left( {u_x^2 + u_y^2} \right),\varphi  + \rho c_s^2\left( {u_x^2 + u_z^2} \right),\varphi  + \rho c_s^2\left( {u_y^2 + u_z^2} \right)} \right]}^{\rm T}},}
\end{array}
\end{equation} 
where $\varphi  = \rho c_s^4\left( {1 - 1.5|{\bf{u}}{|^2}} \right)$ with ${c_s} = \sqrt {{{{c^2}} \mathord{\left/{\vphantom {{{c^2}} 3}} \right.\kern-\nulldelimiterspace} 3}} $ being the lattice sound speed (Here, $c = {{{\delta _x}} \mathord{\left/{\vphantom {{{\delta _x}} {{\delta _t}}}} \right.\kern-\nulldelimiterspace} {{\delta _t}}}$ is the lattice velocity with ${\delta _x}$ denoting the lattice spacing, and it is set to be 1.0 in this work), ${\left| {\bf{u}} \right|^2} = u_x^2 + u_y^2 + u_z^2$, $u_x$, $u_y$, $u_y$ stand for the velocity components in $x$, $y$ and $z$ directions, respectively, $\mathbf{I}$ is a unit matrix and $\boldsymbol{\Lambda}$ is a diagonal matrix given by
\begin{equation}\label{eq7}
	\boldsymbol{\Lambda}=\operatorname{diag}\left(1,1,1,1, s_{e}, s_{v}, s_{v}, s_{v}, s_{v}, s_{v}, s_{q}, s_{q}, s_{q}, s_{q}, s_{q}, s_{q}, s_{\pi}, s_{\pi}, s_{\pi}\right).
\end{equation}
In addition, the variable $\overline{\mathbf{S}}$ is mentioned as a discrete forcing term in the moment space. To ensure good numerical performance while simulating multiphase flow at a relatively large density ratio, the present paper have adopted the forcing scheme proposed by Li et al. \cite{Li_CF_2019}.
\begin{equation}\label{eq8}
\begin{array}{*{20}{c}}
	{\overline {\bf{S}}  = \left[ {0,{F_x},{F_y},{F_z},2{\bf{F}} \cdot {\bf{u}} + \frac{{6\chi {{\left| {{{\bf{F}}_m}} \right|}^2}}}{{{\psi ^2}{\delta _t}\left( {s_e^{ - 1} - 0.5} \right)}},2\left( {2{F_x}{u_x} - {F_y}{u_y} - {F_z}{u_z}} \right),2\left( {{F_y}{u_y} - {F_z}{u_z}} \right),{F_x}{u_y} + {F_y}{u_x},{F_x}{u_z} + {F_z}{u_x},} \right.}\\
	{{F_y}{u_z} + {F_z}{u_y},c_s^2{F_y},c_s^2{F_x},{{\left. {c_s^2{F_z},c_s^2{F_x},c_s^2{F_z},c_s^2{F_y},2c_s^2\left( {{u_x}{F_x} + {u_y}{F_y}} \right),2c_s^2\left( {{u_x}{F_x} + {u_z}{F_z}} \right),2c_s^2\left( {{u_y}{F_y} + {u_z}{F_z}} \right)} \right]}^{\rm{T}}},}
\end{array}
\end{equation}
in which the constant $\chi$ is an adjustment parameter, ${\bf{F}} = {{\bf{F}}_{\rm{m}}} + {{\bf{F}}_{{\rm{ads}}}}$ is the total force exerted on the system with ${{\bf{F}}_{{\rm{m}}}}$ and ${{\bf{F}}_{{\rm{ads}}}}$ representing the liquid-liquid interaction force and the liquid-solid interaction force, respectively, and they are given by \cite{ShanChen_PRE1994,Li_PRE2014_Contact angles}
\begin{equation}\label{eq9}
	{{\bf{F}}_{\rm{m}}}({\bf{x}}) =  - G\psi ({\bf{x}})\sum\limits_\alpha  \omega  \left( {{{\left| {{{\bf{e}}_\alpha }} \right|}^2}} \right)\psi \left( {{\bf{x}} + {{\bf{e}}_\alpha }} \right){{\bf{e}}_\alpha },
\end{equation}
\begin{equation}\label{eq10}
	{{\bf{F}}_{{\rm{ads}}}}({\bf{x}}) =  - {G_w}\psi ({\bf{x}})\sum\limits_\alpha  \omega  \left( {{{\left| {{{\bf{e}}_\alpha }} \right|}^2}} \right)\psi ({\bf{x}})s\left( {{\bf{x}} + {{\bf{e}}_\alpha }} \right){{\bf{e}}_\alpha },
\end{equation}
in which $G$ is the interaction strength, $G_{w}$ is the parameter adopted to control the surface wettability, and $s$ is an indicator function that equals 1.0 for the solid point and 0.0 for the fluid point. $w\left(\left|\mathbf{e}_{\alpha}\right|^{2}\right)$ is the weight factor given by $w_{0}=0$, $w_{1-6}=1/6$ and $w_{7-18}=1/12$, $\psi(\mathbf{x})$ is the interaction potential defined as
\begin{equation}\label{eq11}
	\psi ({\bf{x}}) = \sqrt {\frac{{2\left( {{P_{{\rm{EOS}}}} - \rho c_{\rm{s}}^2} \right)}}{{G{c^2}}}},
\end{equation}
where $P_{\rm{EOS}}$ is the pressure from the piecewise linear equation of state , and it is defined as \cite{Li_ATE_2014,Fei_POF2019}
\begin{equation}\label{eq12}
{P_{EOS}}(\rho ) = \left\{ {\begin{array}{*{20}{l}}
		{\rho {\theta _V}}&{,\rho  \le {\rho _1}}\\
		{{\rho _1}{\theta _V} + \left( {\rho  - {\rho _1}} \right){\theta _M}}&{,{\rho _1} < \rho  \le {\rho _2}}\\
		{{\rho _1}{\theta _V} + \left( {{\rho _2} - {\rho _1}} \right){\theta _M} + \left( {\rho  - {\rho _2}} \right){\theta _L}}&{,\rho  > {\rho _2}}
\end{array}} \right.,
\end{equation}
in which ${\theta _V} = {(\partial {P_{EOS}}/\partial \rho )_V}$, ${\theta _L} = {(\partial {P_{EOS}}/\partial \rho )_L}$ and ${\theta _M} = {(\partial {P_{EOS}}/\partial \rho )_M}$ are the slopes of ${P_{EOS}}(\rho )$ in the vapor-phase region, the liquid-phase region, and the mechanically unstable region, respectively. $\sqrt{\theta_{V}}$ and $\sqrt{\theta_{L}}$ represent the speeds of sound in the vapor and liquid phases, respectively. $\rho_{1}$ and $\rho_{2}$ denote the spinodal points, which are obtained by solving the following two equations \cite{Li_ATE_2014}
\begin{equation}\label{eq13}
	\left(\rho_{1}-\rho_{V}^{\mathrm{e}}\right) \theta_{V}+\left(\rho_{2}-\rho_{1}\right) \theta_{M} +\left(\rho_{L}^{\mathrm{e}}-\rho_{2}\right) \theta_{L}=0,
\end{equation}
\begin{equation}\label{eq14}
	\log \left(\rho_{1} / \rho_{V}^{\mathrm{e}}\right) \theta_{V}+\log \left(\rho_{2} / \rho_{1}\right) \theta_{M}+\log \left(\rho_{L}^{\mathrm{e}} / \rho_{2}\right) \theta_{L}=0,
\end{equation}
in which $\theta_{V}$, $\theta_{L}$, $\theta_{M}$ are model parameters, and can be given as \cite{Fei_POF2019}
\begin{equation}\label{eq15}
	\theta_{V}=c_{s}^{2} / 2, \quad \theta_{L}=c_{s}^{2}, \quad \theta_{M}=-c_{s}^{2} / 40.
\end{equation}
Following the collision step in the moment space, $\mathbf{m}^{*}$ can be reverted back to the discrete velocity space, and then the streaming process can be written as
\begin{equation}\label{eq16}
	f_{\alpha}\left(\mathbf{x}+\mathbf{e}_{\alpha} \delta_{t}, t+\delta_{t}\right)=f_{\alpha}^{*}(\mathbf{x}, t),
\end{equation}
where $\mathbf{f}^{*}=\mathbf{M}^{-1} \mathbf{m}^{*}$, $\mathbf{M}^{-1}$ is the inverse matrix of $\mathbf{M}$. The density and velocity are finally calculated by
\begin{equation}\label{eq17}
	\rho  = \sum\limits_\alpha  {{f_\alpha }} ,\quad \rho {\bf{u}} = \sum\limits_\alpha  {{{\bf{e}}_\alpha }} {f_\alpha } + \frac{{{\delta _t}}}{2}{\bf{F}}.
\end{equation}

\section{Validation}
\label{section4}

\begin{figure}[H]
	\centering
	\includegraphics[width=0.5\textwidth]{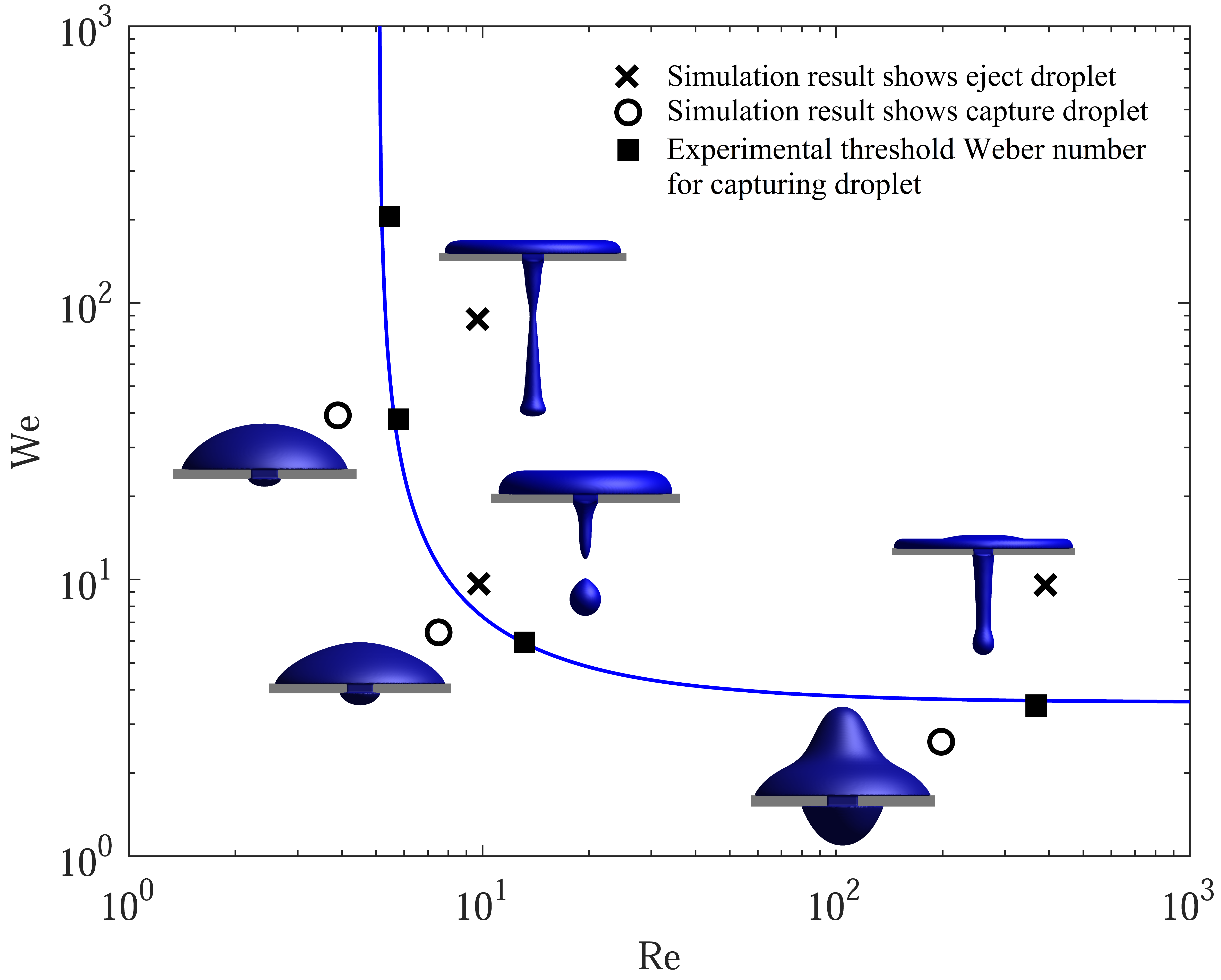}
	\caption{Numerical validation of the present model against the experiment \cite{Lorenceau2003}, in which the solid line ${\mathop{\rm Re}\nolimits} \left( {We - 3.6} \right) = 5.1We$ represents a critical state in which a droplet is able to pass through a perforated plate.}
	\label{fig2}
\end{figure}

In this section, simulations were conducted to analyze the impact of a silicone oil droplet, having a diameter of 3.5 mm, on a stainless steel plate that is 0.25 mm thick and has a pore diameter of 450 um. The results of these simulations were then compared with the experimental data obtained by Lorenceau et al \cite{Lorenceau2003}. The configuration of the problem is similar to that presented in Fig. \ref{fig1} except the wettability of the plate across the thickness is uniform, and the contact angle is approximately $100^\circ$. In our simulations, the grid resolution of the system is fixed at $Nx \times Ny \times Nz = 200 \times 200 \times 250$, which is fine enough to give the grid-independence results, and the liquid density and vapor density are fixed to be 1.0 and 0.00125, respectively. Regarding the boundary conditions, the period boundary is utilized in the horizontal direction, while the no-slip boundary condition is applied in the vertical direction of the computational domain. Experimental evidence shows that a droplet is completely captured by a plate for a given Reynolds number when the Weber number is less than a specific threshold Weber number. It is clear from Fig. \ref{fig2} that the simulation results are consistent with the prior experimental findings \cite{Lorenceau2003}. In particular, it is observed that when the Weber number exceeds the critical threshold at a given Reynolds number, the liquid is expelled from the plate surface and breaks down into smaller droplets. Conversely, if the Weber number falls below the critical threshold, a liquid finger briefly emerges and recedes before the entire liquid is captured by the surface.

\section{Results and discussion}\label{section5}

Below, we present and discuss the numerical results of a droplet impacting a hydrophobic plate with a wettability-patterned orifice. The influences of wettability difference ($\Delta \theta  = {\theta _1} - {\theta _2}$), the Weber number ($We$)  as well as the pore size ($r/R$) are all investigated in detail. Unless otherwise stated, in the simulations, the computational mesh adopted here is the same as that used in numerical validation, while the radius of the droplet is set to be 20.0 lattice units. In addition, the contact angle of the top plate in what follows is fixed at $\theta_1=150^\circ$. The density ratio between liquid and vapor phase is set to be 800.0, i.e., $\rho_L/\rho_V=800.0$, and the kinetic viscosity ratio is fixed at 10.0, i.e., ${{{\nu _V}} \mathord{\left/{\vphantom {{{\nu _V}} {{\nu _L} = 10.0}}} \right.\kern-\nulldelimiterspace} {{\nu _L} = 10.0}}$. In the diagonal matrix, ${s_e} = 0.8,{s_q} = {s_\pi } = 1.2,{s_\nu } = \tau _\nu ^{ - 1}$, and ${\tau _\nu }$ is the local relaxation time. According to a scientific study \cite{Li_PRE2013}, it has been observed that the ratio between the density of water and air, and the ratio between the kinematic viscosity of water and air, are quite similar to the two values mentioned. Further, the dimensionless time adopted in what follows is defined as ${t^*} = {t \mathord{\left/{\vphantom {t {\sqrt {{{{\rho _L}{R^3}} \mathord{\left/{\vphantom {{{\rho _L}{R^3}}\sigma}}\right.\kern-\nulldelimiterspace} \sigma }} }}} \right.\kern-\nulldelimiterspace} {\sqrt {{{{\rho _L}{R^3}} \mathord{\left/{\vphantom {{{\rho _L}{R^3}}\sigma}}\right.\kern-\nulldelimiterspace} \sigma }} }}$. For the boundary condition used here, the no-slip boundary condition is applied in the vertical direction, while the periodic boundary condition is employed in the horizontal direction.

\subsection{Effect of wettability difference}
\label{section5.1}

\begin{figure}
	\centering
	\includegraphics[width=0.8\textwidth]{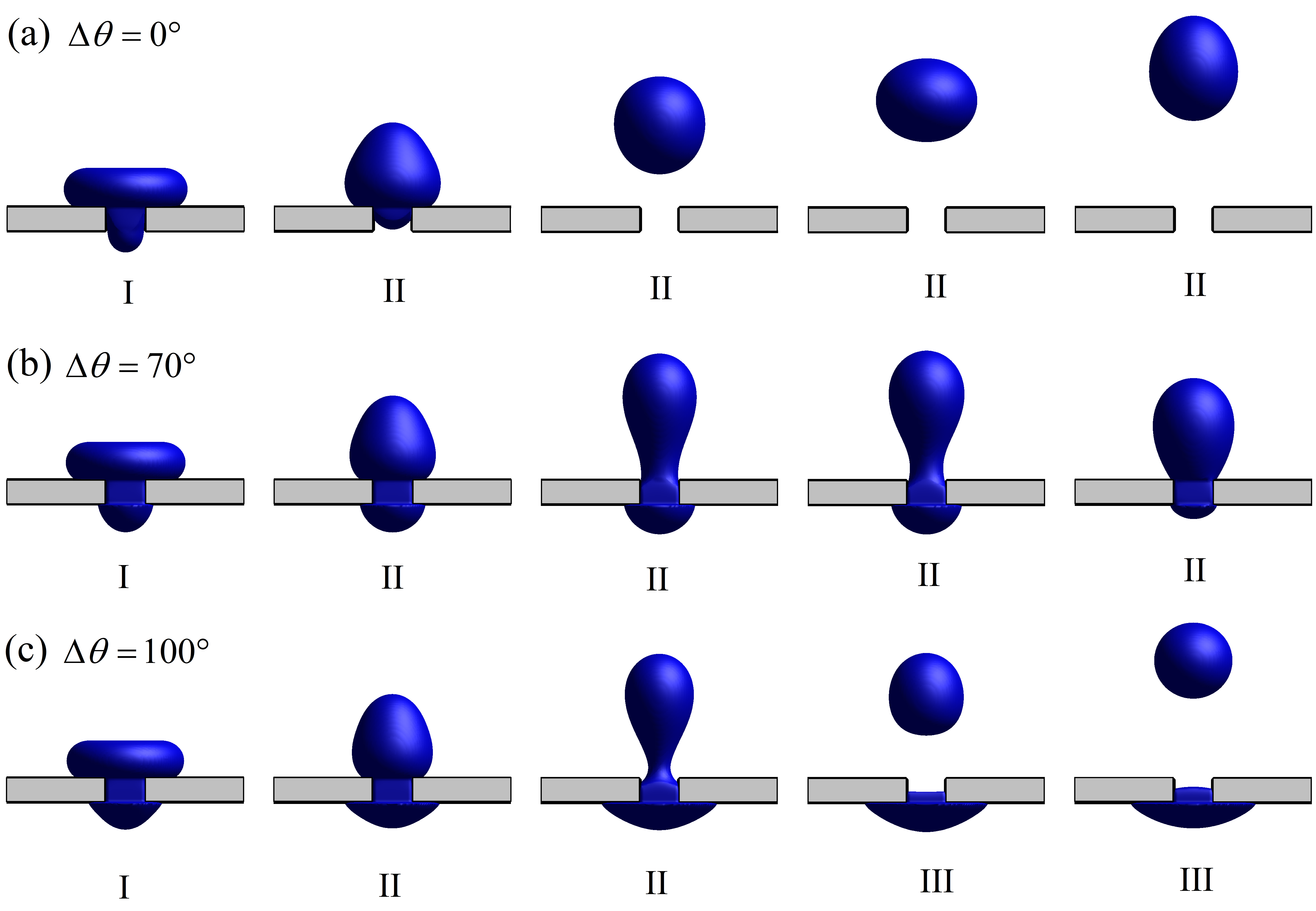}	
	\caption{The morphologies of droplet impacting on the perforated plate for  $\Delta \theta = 0^\circ$ (a), $\Delta \theta =70^\circ$ (b) and $\Delta \theta = 100^\circ$ (c) at $We=10.0$, $r/R=0.4$, and $h/R=0.5$, in which I, II and III represent the spreading phase, rebounding phase, and splitting phase, respectively.}
	\label{fig3}
\end{figure}

We first investigate the effect of the wettability difference of the pore. In the simulations,  the contact angle of the bottom plate (i.e., ${\theta_2}$ ) varies from $50^\circ$ to $150^\circ$. The length ratio between pore and droplet radii is set to be  $r/R=0.4$, while the ratio between plate thickness and droplet radius is fixed at $h/R=0.5$. In addition, the values of the Weber number, capillary number, and Ohnesorge number are fixed at $We=10.0$, $Ca=0.125$, and $Oh=0.016$, respectively. Fig. \ref{fig3} illustrates the morphologies of the droplet impinging upon a plat with a wettability-patterned pore, in which the evolutionary phase is also presented. It is found that when the pore has homogeneous wettability, the droplet usually first spreads under the influence of inertial force and then rebounds above the plate, which is similar to that observed in previous work \cite{Wang_PRF_2020,Lorenceau2003}. However, when the pore wettability is heterogeneous, the droplet impact behavior is largely different. Specifically, owing to the hydrophilic nature of the bottom plate, during the spreading stage, a small amount of liquid droplet adheres to the bottom surface. In the subsequent rebounding stage, due to the influence of the unbalanced Young's force, the upward rebounding of the liquid droplet is hindered to some extent. As a result, a droplet can either be captured by the plate or split into two daughter droplets (see Fig. \ref{fig3}(b) and Fig. \ref{fig3}(c)). 

To better comprehend how droplets interact with a hydrophobic surface featuring a wettability-patterned pore, the mechanisms behind each phase are discussed in detail in what follows. As shown in Fig. \ref{fig3}, we can see that the droplet impacts the center of the perforated plate surface vertically during the spreading stage, driven by the downward inertial force. Finally, a liquid finger is formed within the pore, accompanied by a minor quantity of the droplet adhering to the bottom hydrophilic plate. However, the droplet trapped at the top hydrophobic plate continues to spread until the motion of the contact line terminates, at which point the inertial force, the viscous force, and the capillary force reach an equilibrium state. After that, owing to the surface tension effect, the droplet starts to retract towards the center and enters the rebounding phase. Different from the homogeneous pore (see Fig. \ref{fig3}(a)), for which the whole droplet leaves the plate at the rebounding phase. Due to the difference in wettability across the thickness of the heterogeneous pore, an unbalanced net force is generated in the downward direction, which causes a part of the droplet to migrate directionally towards the more wetting region. Eventually, the droplet may split into two daughter droplets depending on the wettability difference. Comparing Fig. \ref{fig3}(b) and Fig. \ref{fig3}(c), one can conclude that an increase in the pore wettability difference results in an increase in the amount of the adhering droplet.  

\begin{figure}[H]
	\centering
	\subfigure[]{\label{fig4a}
		\includegraphics[width=0.46\textwidth]{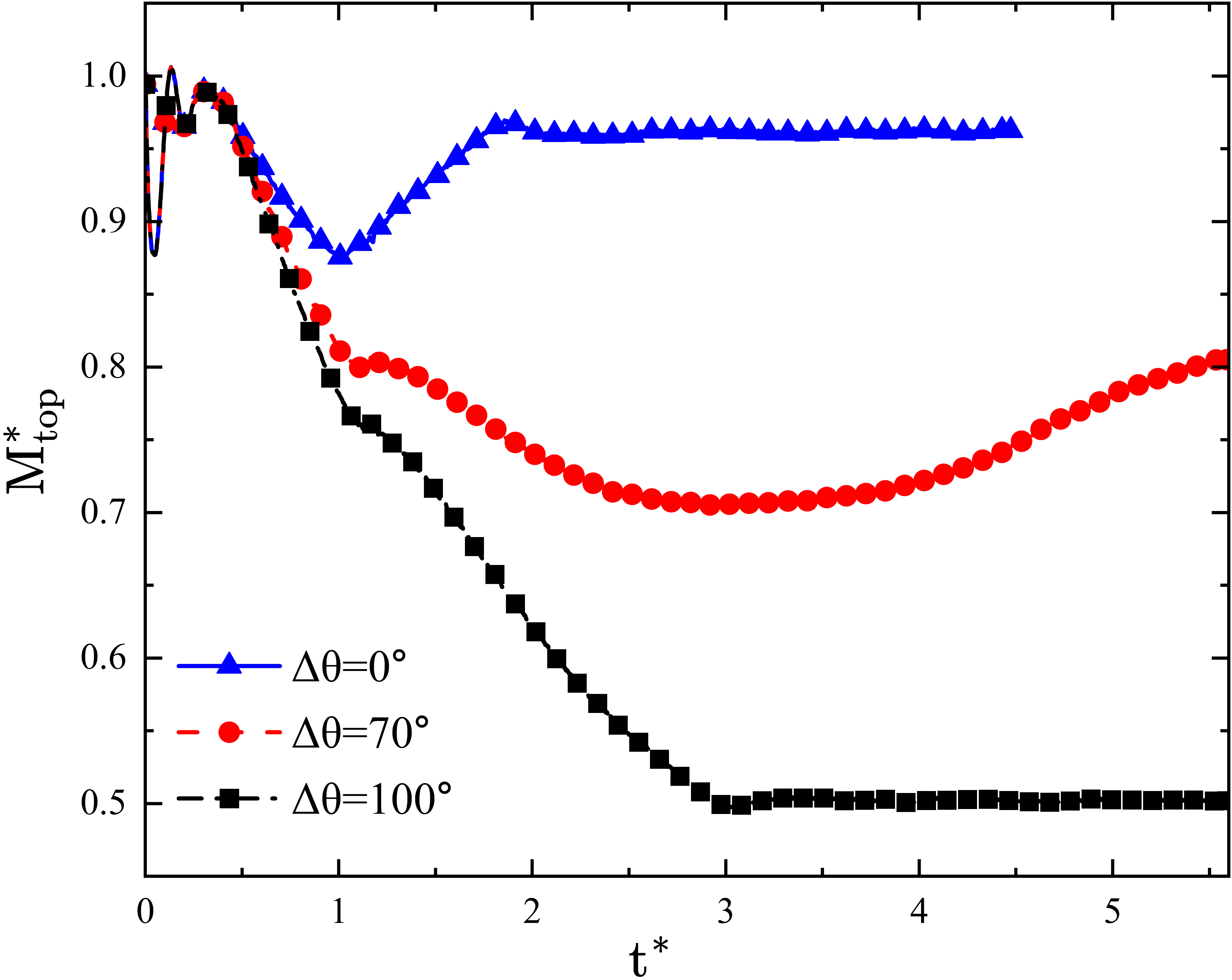}
	}
	\subfigure[]{\label{fig4b}
		\includegraphics[width=0.46\textwidth]{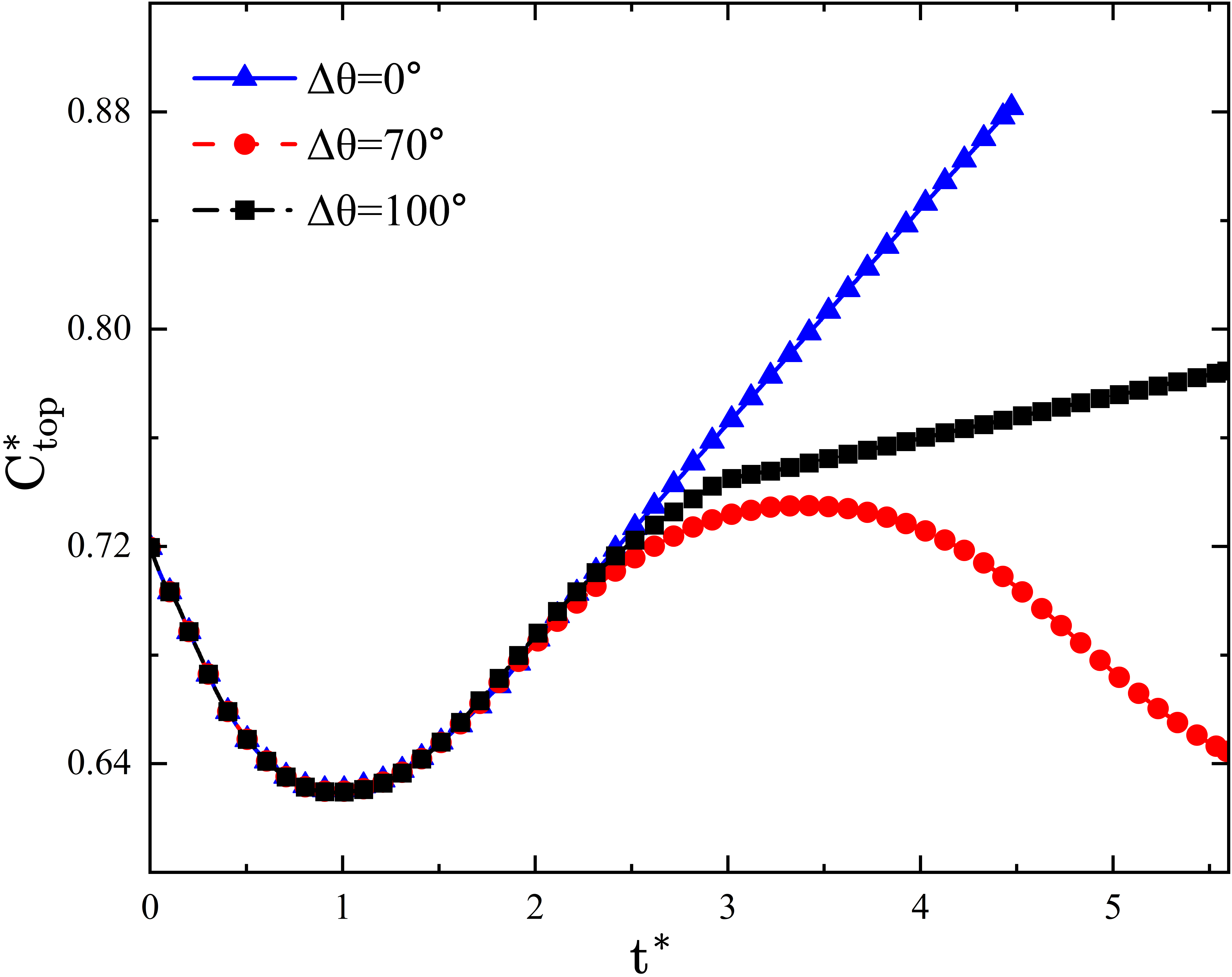}
	}
	\label{fig4}
	\caption{The dimensionless mass (a) and mass center (b) evolution of droplets above the plate for various wettability differences at $We=10.0$, $r/R=0.4$, and $h/R=0.5$.}
\end{figure}	
	
Another point worth mentioning is that for the heterogeneous pore considered here, the droplet may encounter an oscillation behavior. To this end,  the residual mass ${\rm{M}}_{{\rm{top}}}^{\rm{*}}$ as well as its mass center ${\rm{C}}_{{\rm{top}}}^{\rm{*}}$ above the plate are recorded (Noting that since the penetrated mass below the plate is too little, the daughter-droplet in this position is not considered here), and the mass center ${\rm{C}}_{{\rm{top}}}$ is calculated by
\begin{equation}\label{eq18}
	{\rm{C}}_{{\rm{top}}} = \frac{{\sum\limits_{\rho  < {{\left( {{\rho _L} + {\rho _V}} \right)} \mathord{\left/
	{\vphantom {{\left( {{\rho _L} + {\rho _V}} \right)} 2}} \right.
	\kern-\nulldelimiterspace} 2}} {C\left( {{\bf{x}},t} \right)\rho \left( {{\bf{x}},t} \right)} }}{{\sum\limits_{\rho  < {{\left( {{\rho _L} + {\rho _V}} \right)} \mathord{\left/
	{\vphantom {{\left( {{\rho _L} + {\rho _V}} \right)} 2}} \right.
	\kern-\nulldelimiterspace} 2}} {\rho \left( {{\bf{x}},t} \right)} }},
\end{equation}	
where ${{\left( {{\rho _L} + {\rho _V}} \right)} \mathord{\left/{\vphantom {{\left( {{\rho _L} + {\rho _V}} \right)} 2}} \right.\kern-\nulldelimiterspace} 2}$ denotes the phase interface and $C$ represents the vertical coordinate. Fig. \ref{fig4a} and Fig. \ref{fig4b} illustrate how the residual mass and the mass center are affected when impacting the plat with different wettability differences, in which ${\rm{M}}_{{\rm{top}}}^{\rm{*}}$ and ${\rm{C}}_{{\rm{top}}}^{\rm{*}}$ are normalized by the droplet initial mass and the grid number in $z-$direction, respectively. Based on this figure, one can find that at the initial spreading stage, the mass centers for different wettability differences are all firstly decrease until it reaches a minimum value. After that, the mass center obtained for the homogeneous pore continues to rise until it reaches the top of the calculated region in the absence of gravitational effects. For the heterogeneous pore with $\Delta \theta  = 100^\circ $, although the mass center of the droplet is also increased when it leaves the surface, there exists a turning point at which the droplet splits into two daughter ones, in the curve, and the mass center is found to increase with a relatively smaller rate after that. Different from the above two cases, as far as the heterogeneous pore with $\Delta \theta  = 70^\circ $ is concerned, due to the competition between the dynamic pressure ($0.5{\rho _L}{U^2}$) and the unbalanced Young's force in the system, the mass center of the droplet in such a case exhibits an oscillating behavior, as shown in Fig. \ref{fig4b}. The above analysis clearly indicates that the presence of heterogeneous pore has a significant influence on the droplet impact behavior, it is expected that the residual mass above the plate depends on the corresponding configuration. As shown in Fig. \ref{fig4a}, one can clearly see that for a given dimensionless time, the residual mass is decreased with the increase of the wettability difference, which is consistent with its dynamic process discussed above.

\begin{figure}[H]
	\centering
	\subfigure[]{\label{fig5a}
		\includegraphics[width=0.46\textwidth]{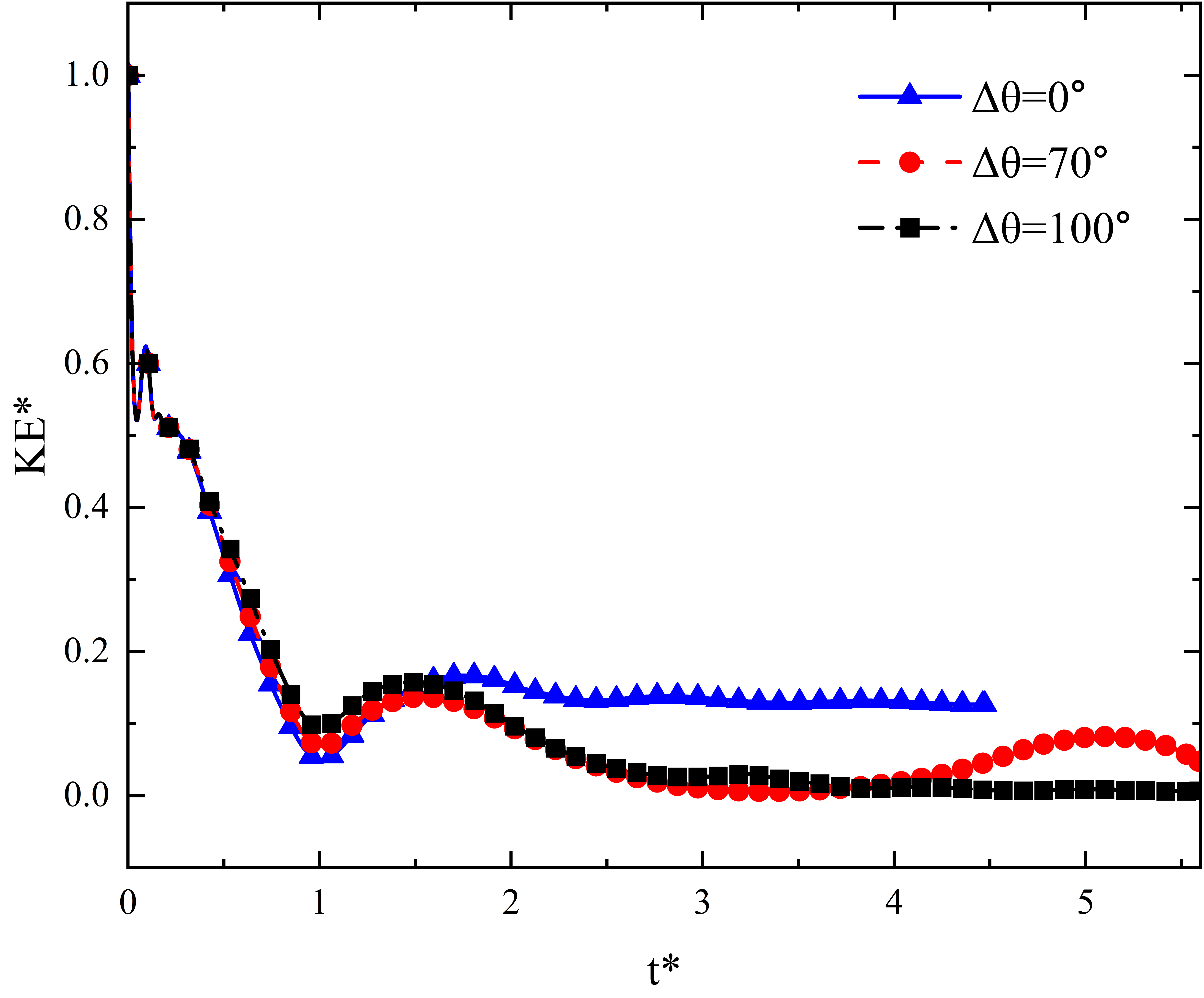}
	}
	\subfigure[]{\label{fig5b}
		\includegraphics[width=0.46\textwidth]{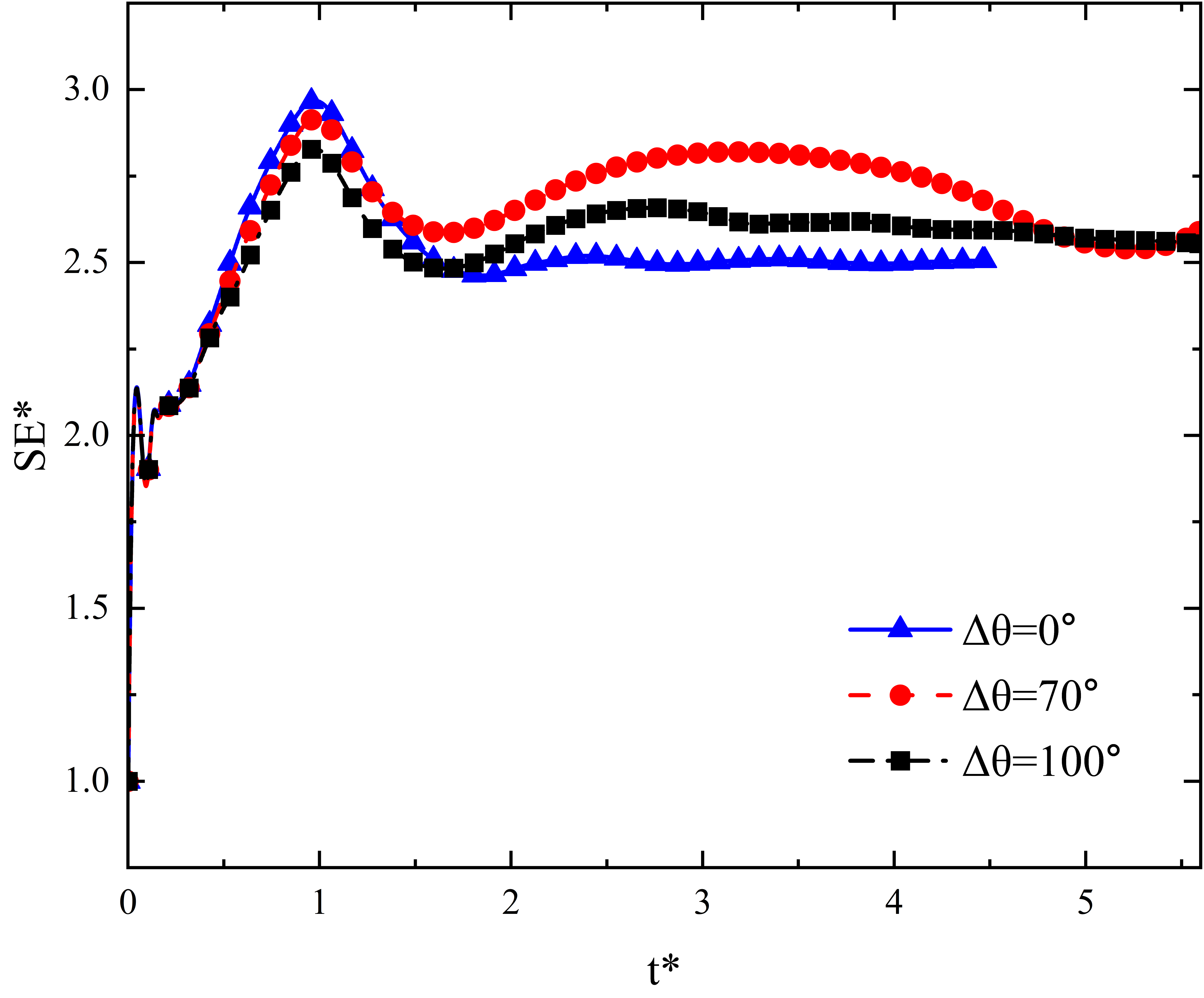}
	}
	\caption{Temporal evolution of the dimensionless kinetic energy (a) and surface energy (b) for various  wettability differences at $We=10.0$, $r/R=0.4$, and $h/R=0.5$.}
	\label{fig5}
\end{figure}

To further quantitatively investigative the droplet impact mechanism, we also conduct a detailed energy analysis on the droplet impact dynamics, and the related energies include the kinetic energy ($KE$), the surface energy ($SE$), and the viscous dissipation energy ($VD$), which are given by  
\begin{equation} \label{eq19}
	\begin{array}{ll}
		KE=\iiint 0.5 \rho \left( {u_x^2 + u_y^2 + u_z^2} \right)d\Omega , \quad SE=\sigma S, \quad \mathrm{VD}=\int_{0}^{t} \int_{0}^{\Omega} \phi d \Omega \mathrm{d} t,
	\end{array}
\end{equation}
where $\Omega $ denotes fluid domain (liquid and gas), $S$ is the area of the droplet interface and $U$ represents the droplet velocity. In addition, $\int_0^\Omega  \phi  d\Omega $ is the viscous dissipation rate ($VDR$) with $\phi$ being the dissipation function, which quantifies the local volumetric viscous dissipation rate, and it is defined as \cite{He_POF_2019}
\begin{equation} \label{eq20}
	\phi  = {\mu}\left[ {2{{\left( {\frac{{\partial {u_x}}}{{\partial x}}} \right)}^2} + 2{{\left( {\frac{{\partial {u_y}}}{{\partial y}}} \right)}^2} + 2{{\left( {\frac{{\partial {u_z}}}{{\partial z}}} \right)}^2} + {{\left( {\frac{{\partial {u_x}}}{{\partial y}} + \frac{{\partial {u_y}}}{{\partial x}}} \right)}^2} + {{\left( {\frac{{\partial {u_y}}}{{\partial z}} + \frac{{\partial {u_z}}}{{\partial y}}} \right)}^2} + {{\left( {\frac{{\partial {u_z}}}{{\partial x}} + \frac{{\partial {u_x}}}{{\partial z}}} \right)}^2}} \right],
\end{equation}
in which the gradient of velocity component (${{\partial {u_x}} \mathord{\left/
		{\vphantom {{\partial {u_x}} {\partial x}}} \right.
		\kern-\nulldelimiterspace} {\partial x}},{{\partial {u_x}} \mathord{\left/
		{\vphantom {{\partial {u_x}} {\partial y}}} \right.
		\kern-\nulldelimiterspace} {\partial y}},{{\partial {u_x}} \mathord{\left/
		{\vphantom {{\partial {u_x}} {\partial z}}} \right.
		\kern-\nulldelimiterspace} {\partial z}},{{\partial {u_y}} \mathord{\left/
		{\vphantom {{\partial {u_y}} {\partial x}}} \right.
		\kern-\nulldelimiterspace} {\partial x}},{{\partial {u_y}} \mathord{\left/
		{\vphantom {{\partial {u_y}} {\partial y}}} \right.
		\kern-\nulldelimiterspace} {\partial y}},{{\partial {u_y}} \mathord{\left/
		{\vphantom {{\partial {u_y}} {\partial z}}} \right.
		\kern-\nulldelimiterspace} {\partial z}},{{\partial {u_z}} \mathord{\left/
		{\vphantom {{\partial {u_z}} {\partial x}}} \right.
		\kern-\nulldelimiterspace} {\partial x}},{{\partial {u_z}} \mathord{\left/
		{\vphantom {{\partial {u_z}} {\partial y}}} \right.
		\kern-\nulldelimiterspace} {\partial y}},{{\partial {u_z}} \mathord{\left/
		{\vphantom {{\partial {u_z}} {\partial z}}} \right.
\kern-\nulldelimiterspace} {\partial z}}$) is calculated by using a second-order isotropic central schemes \cite{Liang_PRE_2016}. Take ${{\partial {u_x}} \mathord{\left/{\vphantom {{\partial {u_x}} {\partial x}}} \right.\kern-\nulldelimiterspace} {\partial x}}$ as an example, 
\begin{equation}\label{eq21}
\frac{{\partial {u_x}}}{{\partial x}} = \sum\limits_{i \ne 0} {\frac{{{\varpi _i}{{\bf{e}}_{i,x}}{u_x}\left( {{\bf{x}} + {{\bf{e}}_{i,x}}\delta x} \right)}}{{c_s^2\delta t}}} ,
\end{equation}
where ${{\varpi _i}}$ is the weighting coefficient, and its value in D3Q19 lattice model is given by ${\varpi _{1 - 6}} = {1 \mathord{\left/{\vphantom {1 {18}}} \right.\kern-\nulldelimiterspace} {18}},{\varpi _{7 - 18}} = {1 \mathord{\left/
{\vphantom {1 {36}}} \right.	\kern-\nulldelimiterspace} {36}}.$ Similar to the previous study \cite{Finotello_POF_2017}, the current work also considers the viscous dissipation in the gas phase, and thus the viscosity ${\mu}$ appeared in Eq. (\ref{eq20}) is actually a local value. In such a case, from the point of view of energy conservation, the total energy $TE$ is the amount of the initial kinetic energy $KE_{init}$ and the initial surface energy $SE_{init}$ of the droplet, and it equals the sum of the kinetic, surface and viscous dissipated energy at a given time, 
\begin{equation}\label{eq22}
	TE = KE_{init} + SE_{init} = KE(t) + SE(t) + VD(t).
\end{equation}

\begin{figure}[h]
	\centering
	\includegraphics[width=0.46\textwidth]{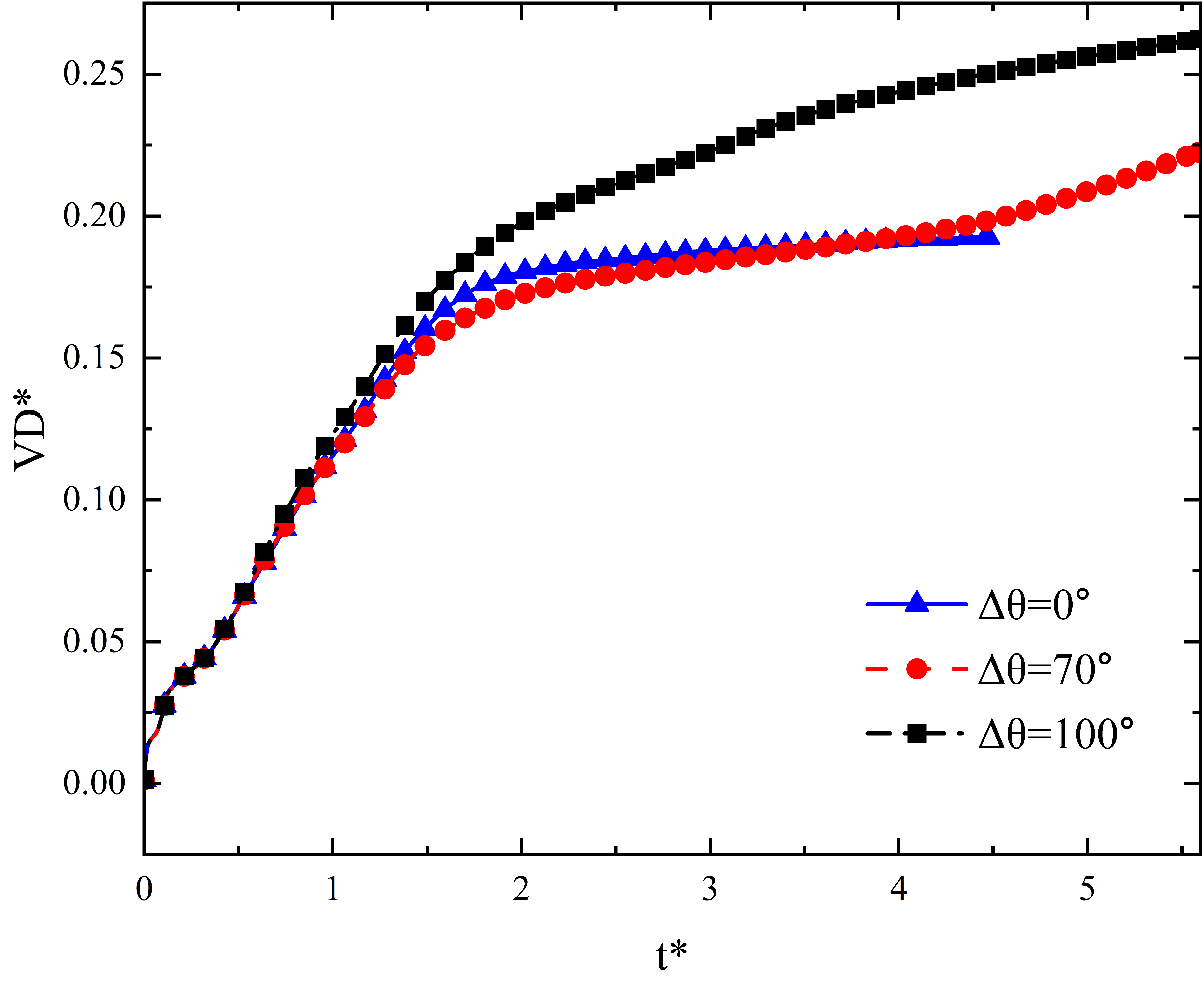}	
	\caption{Temporal evolution of the dimensionless dissipated energy for  various  wettability differences at $We=10.0$, $r/R=0.4$, and $h/R=0.5$.}
	\label{fig6}
\end{figure}

Fig. \ref{fig5a} and \ref{fig5b} present the evolution of the kinetic energy and the surface energy, in which the dimensionless energy $S{E^*}$and $K{E^*}$ are normalized by their initial values. It is observed that at the spreading phase ($0.0 < {t^*} < 1.0$), the kinetic energy significantly decreases, and the decreased kinetic energy can be divided into two parts: one part undergoes conversion into surface energy, while another part dissipates through viscous dissipation energy. As a result, the surface energy and the dissipated energy in this state are both increased rapidly for all cases (see Fig. \ref{fig5b} and Fig. \ref{fig6}). In particular, since the maximum spreading area increases in wettability difference, the surface energy at the final spreading stage (${t^*} > 1.5$) observed at $\Delta \theta  = 70^\circ$ is larger than the other two cases, and this tendency always holds thereafter. Additionally, at the beginning of the rebounding phase ($1.0 < {t^*} < 2.0$), it is interesting to note that the kinetic energies in three cases all firstly increase ($1.0 < {t^*} < 1.5$) and then decrease (${t^*} > 1.5$). The enhancement of kinetic energy in these three cases is similar, and it is mainly caused by the fact that a part of the surface energy during the spreading stage is converted into kinetic energy. However, the mechanisms behind the reduction of kinetic energy for different wettability differences are largely different. The case where $\Delta \theta = 0^\circ$ is due to the energy dissipation caused by the droplet's deformation during departure from the surface (see Fig. \ref{fig3}(a)), while for $\Delta \theta  = 70^\circ$ and $\Delta \theta  = 100^\circ$, apart from the above viscous energy dissipation, a significant portion of the droplet's energy is expended in pulling the liquid towards a region of greater wetting due to the unbalanced Young's force. As a consequence, it is established that as $1.5 < {t^*} < 2.5$, the droplet kinetic energy significantly decreases for heterogeneous pore, and this phenomenon is more distinct for a larger wettability difference. Further, at the final stage of the droplet impacting, owing to the droplet oscillation behavior at $\Delta \theta  = 70^\circ$, the variation of the surface energy in such a case also exhibits an oscillation behavior, while it is nearly constant for  $\Delta \theta  = 0^\circ$. Moreover, Comparing the curves in Fig. \ref{fig5b}, one can deduce that the maximum droplet surface area for the case of $\Delta \theta  = 70^\circ$ is always larger than that for the case of $\Delta \theta  = 100^\circ$. Finally, as depicted in Fig. \ref{fig6}, it is found that a larger wettability difference usually leads to a higher viscous dissipation rate, which suggests that the heterogeneous plat will lead to a higher dissipated energy. 

\subsection{Effect of Weber number}
\label{section5.2}

\begin{figure}[H]
	\centering
	\includegraphics[width=0.8\textwidth]{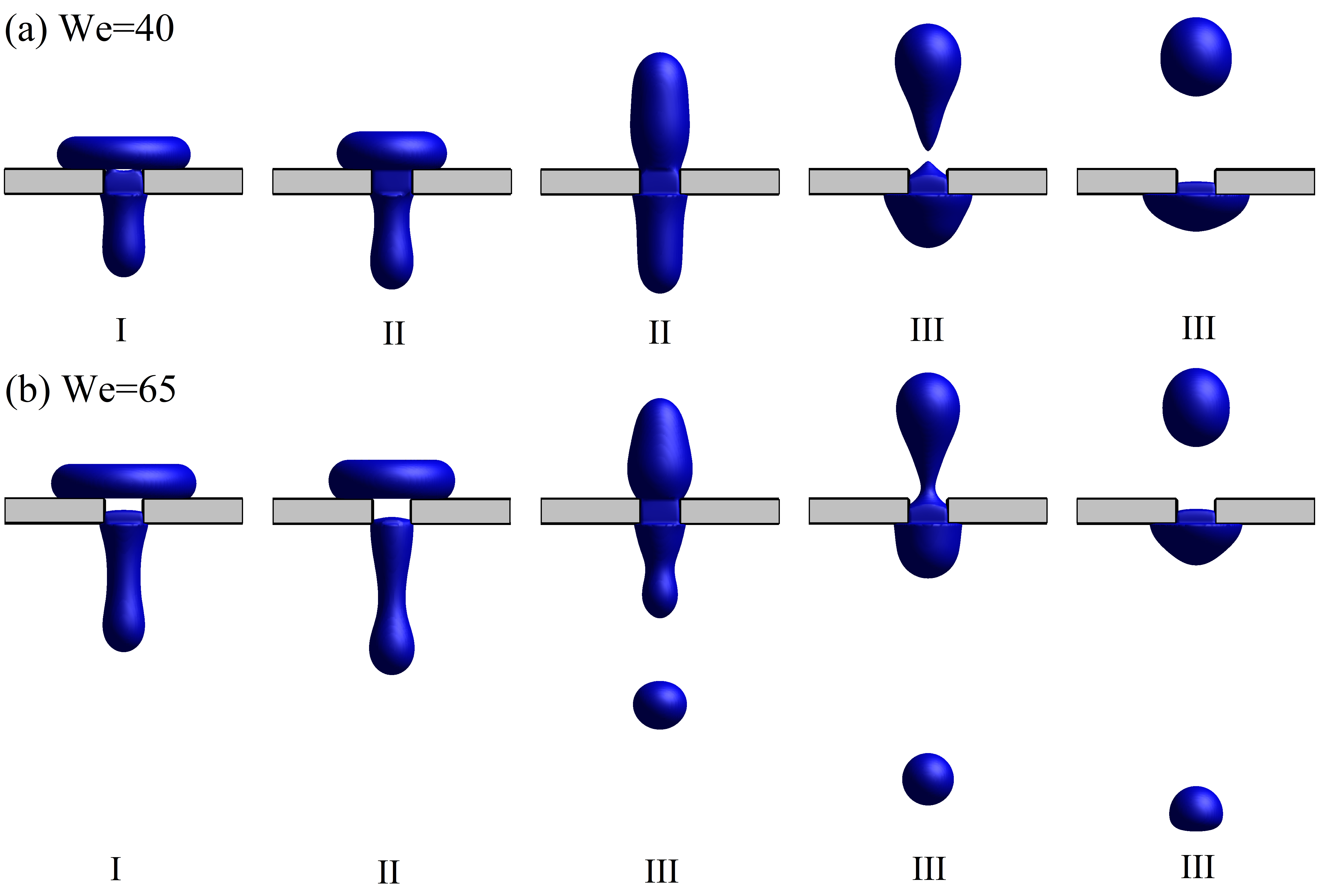}
	\caption{The morphologies of droplet impacting on the perforated plate for $We=40.0$ and $We=65.0$ at $\Delta \theta  = 70^\circ $, $r/R=0.4$, and $h/R=0.5$, in which I, II and III represent the spreading phase, rebounding phase, and the splitting phase.}
	\label{fig7}
\end{figure}

\begin{figure}[H]
	\centering
	\includegraphics[width=0.5\textwidth]{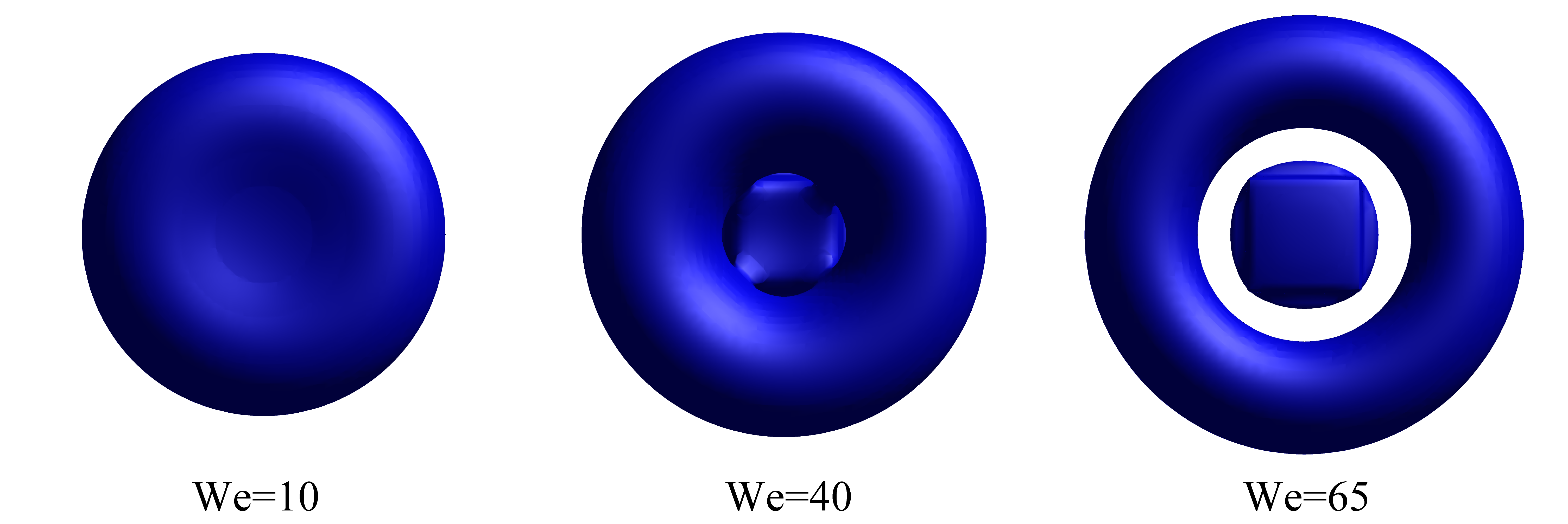}
	\caption{Top view of the droplet morphologies for $We=10.0$ (left), $We=40.0$ (middle) and $We=65.0$ (right) during the spreading phase, in which $\Delta \theta  = 70^\circ $, $r/R=0.4$, $h/R=0.5$, and ${{\rm{t}}^{\rm{*}}}{\rm{ = 0}}{\rm{.9}}$.}
	\label{fig8}
\end{figure}

We now turn to investigate the influence of the Weber number on the droplet impact behavior on a hydrophobic plate with a wettability-patterned pore. In simulations, the initial velocity of the droplet is varied to adjust Weber number, which is always adopted in the literature. The wettability difference across the thickness is set to be $\Delta \theta  = 70^\circ$. The length ratio between pore and droplet radii is set to be  $r/R=0.4$, while the ratio between plate thickness and droplet radius is fixed at $h/R=0.5$. Fig. \ref{fig7} illustrates the droplet evolution process for $We=40.0$ and $We=65.0$. During the initial spreading phase, a liquid finger develops at the bottom of the plate, and its length increases as the Weber number increases. Particularly, it is interesting to note that when $We=65.0$, after droplet impacting, under the combined action of downward unbalanced Young's force and inertial force, the dewetting phenomenon occurs inside the pore (see Fig. \ref{fig7}). Another point that needs to be mentioned is that a dry surface tends to expose near the pore at the top plate after droplet impacting, and this phenomenon is even more pronounced for a comparatively higher Weber number, as illustrated in Fig. \ref{fig8}. As time goes on, the above-mentioned liquid finger beneath the plate continuously develops. In the case of $We=65.0$, owing to the extensive radial dispersion of the liquid on the upper plate, the dynamic pressure ($0.5{\rho _L}{U^2}$) directly above the orifice diminishes, rendering it incapable of surpassing the opposing contact force. A sub-volume of the droplet breaks away from the pore while a small fraction of the droplet remains pinned and hanging at the bottom of the plate.  Also, the liquid inside the pore in such a case wets the dry surface. In the scenario where $We=40.0$, approximately fifty percent of the droplet eventually comes to rest and establishes equilibrium contact below the plate. As for the remaining droplet above the plate, although the impact behaviors in these two cases are different from that observed at $We=10.0$ (see Fig. \ref{fig3}(b)), they all undergo the rebounding and splitting phases, which suggests that the variation of the Weber number has a significant effect on droplet impact behavior. 
	
\begin{figure}[H]
	\centering
	\subfigure[]{\label{fig9a}
		\includegraphics[width=0.46\textwidth]{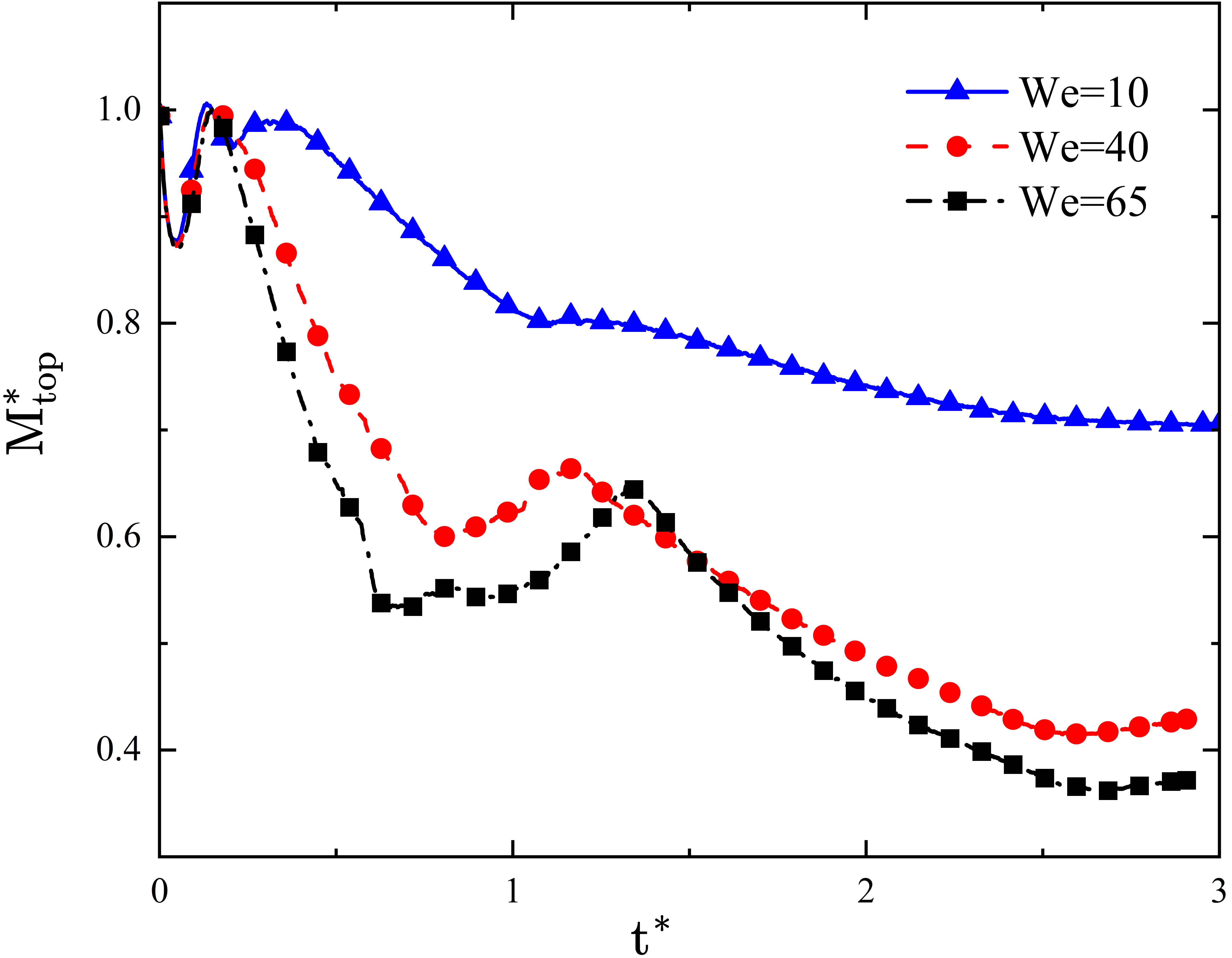}
	}
	\subfigure[]{\label{fig9b}
		\includegraphics[width=0.46\textwidth]{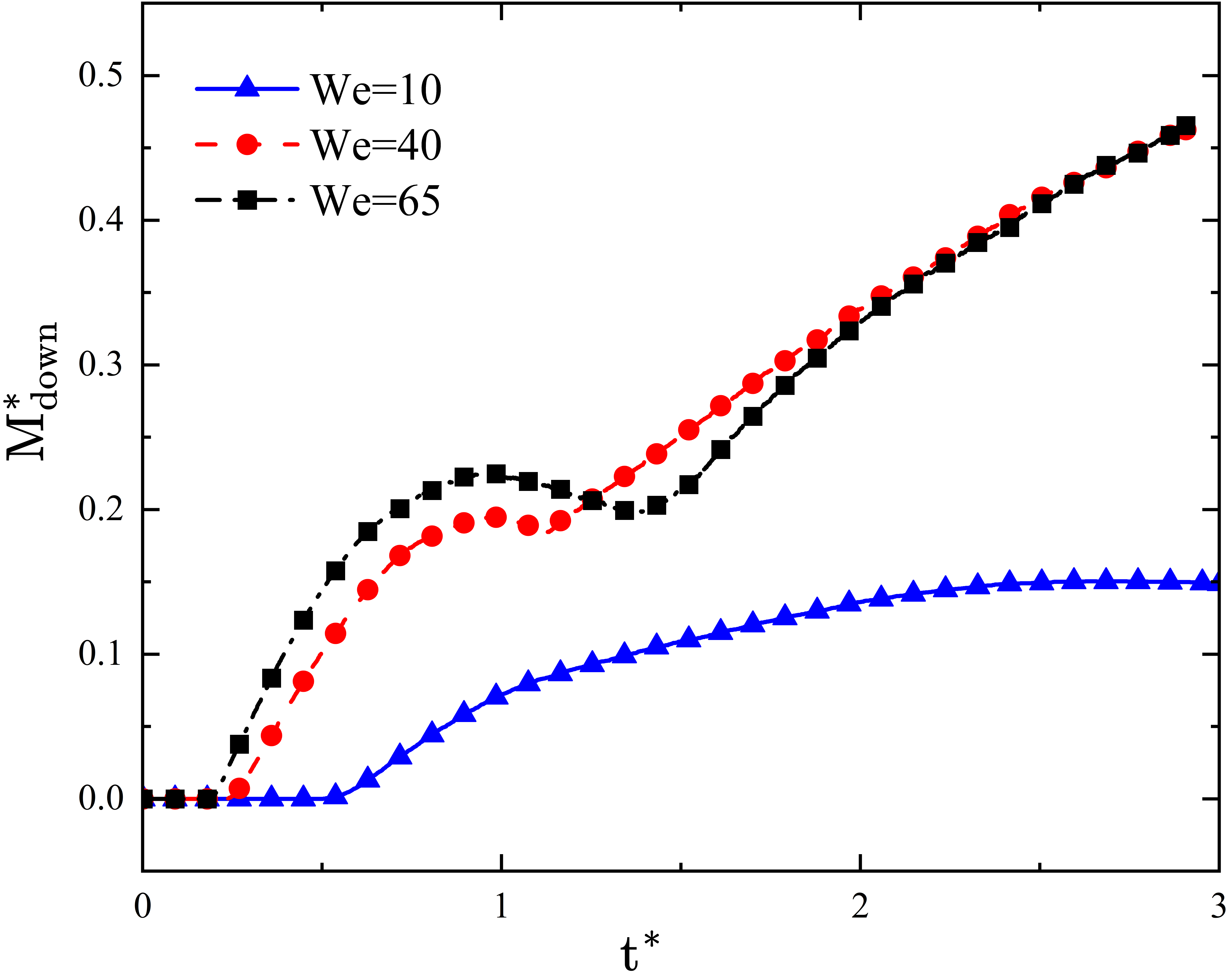}
	}
	\caption{The dimensionless mass evolution of droplets above (a) and below (b) the plate for various Weber numbers at $\Delta \theta  = 70^\circ$, $r/R=0.4$, and $h/R=0.5$.}
	\label{fig9}
\end{figure}

\begin{figure}[H]
	\centering
	\subfigure[]{\label{fig10a}
		\includegraphics[width=0.46\textwidth]{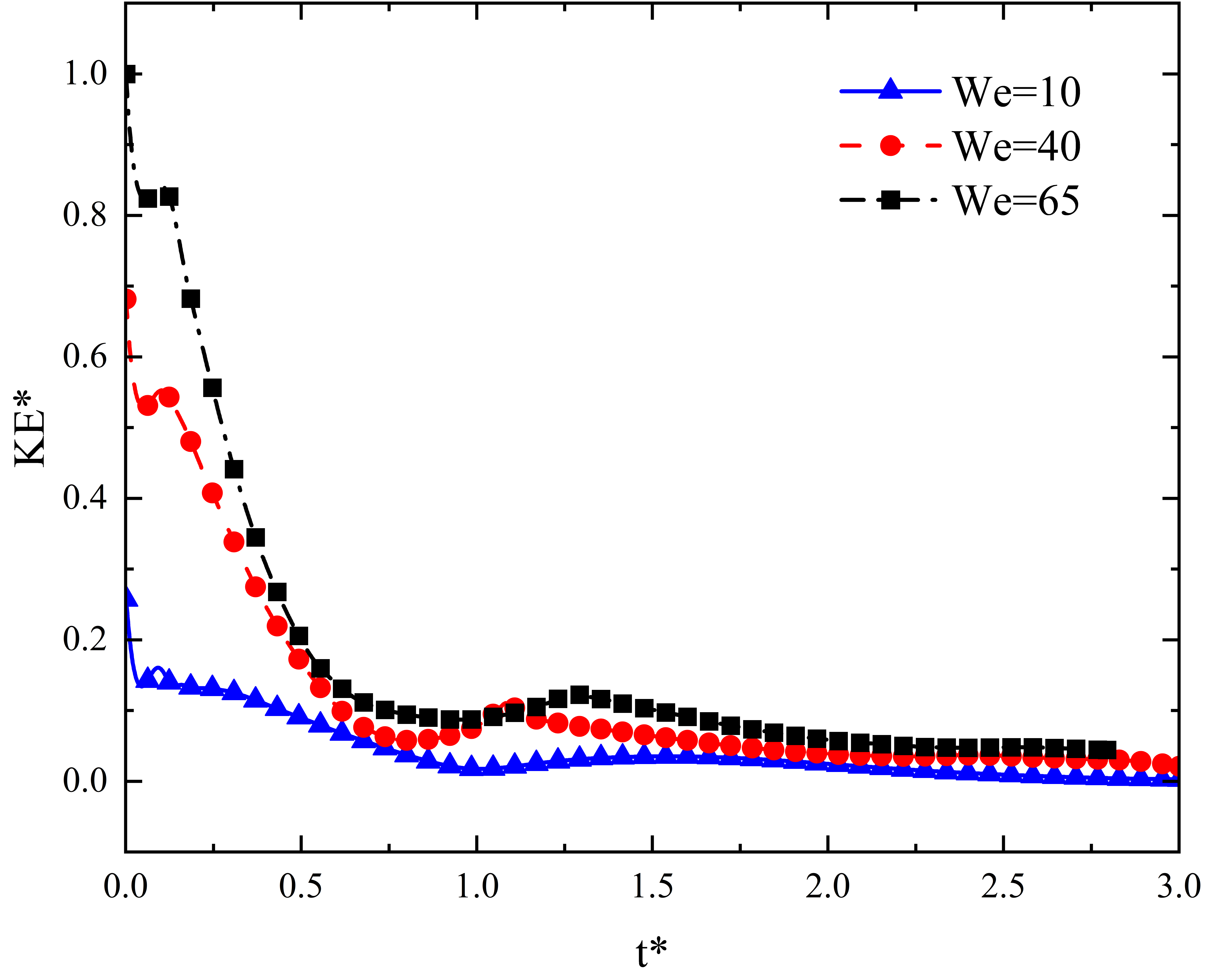}
	}
	\subfigure[]{\label{fig10b}
		\includegraphics[width=0.46\textwidth]{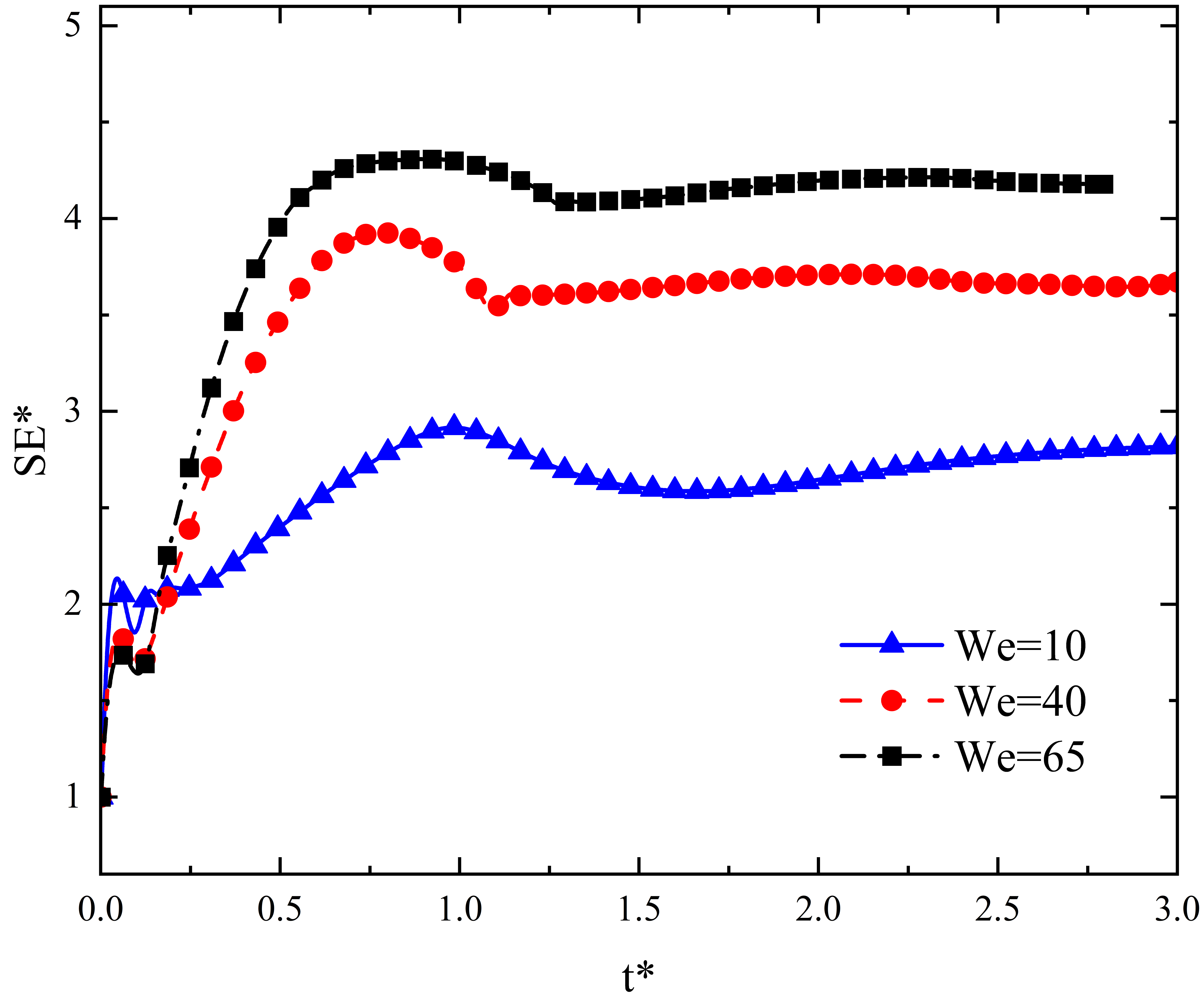}
	}
	\caption{Temporal evolution of the dimensionless kinetic energy (a) and surface energy (b) for various Weber numbers at $\Delta \theta  = 70^\circ$, $r/R=0.4$, and $h/R=0.5$.}
	\label{fig10}
\end{figure}

\begin{figure}[H]
	\centering
	\includegraphics[width=0.46\textwidth]{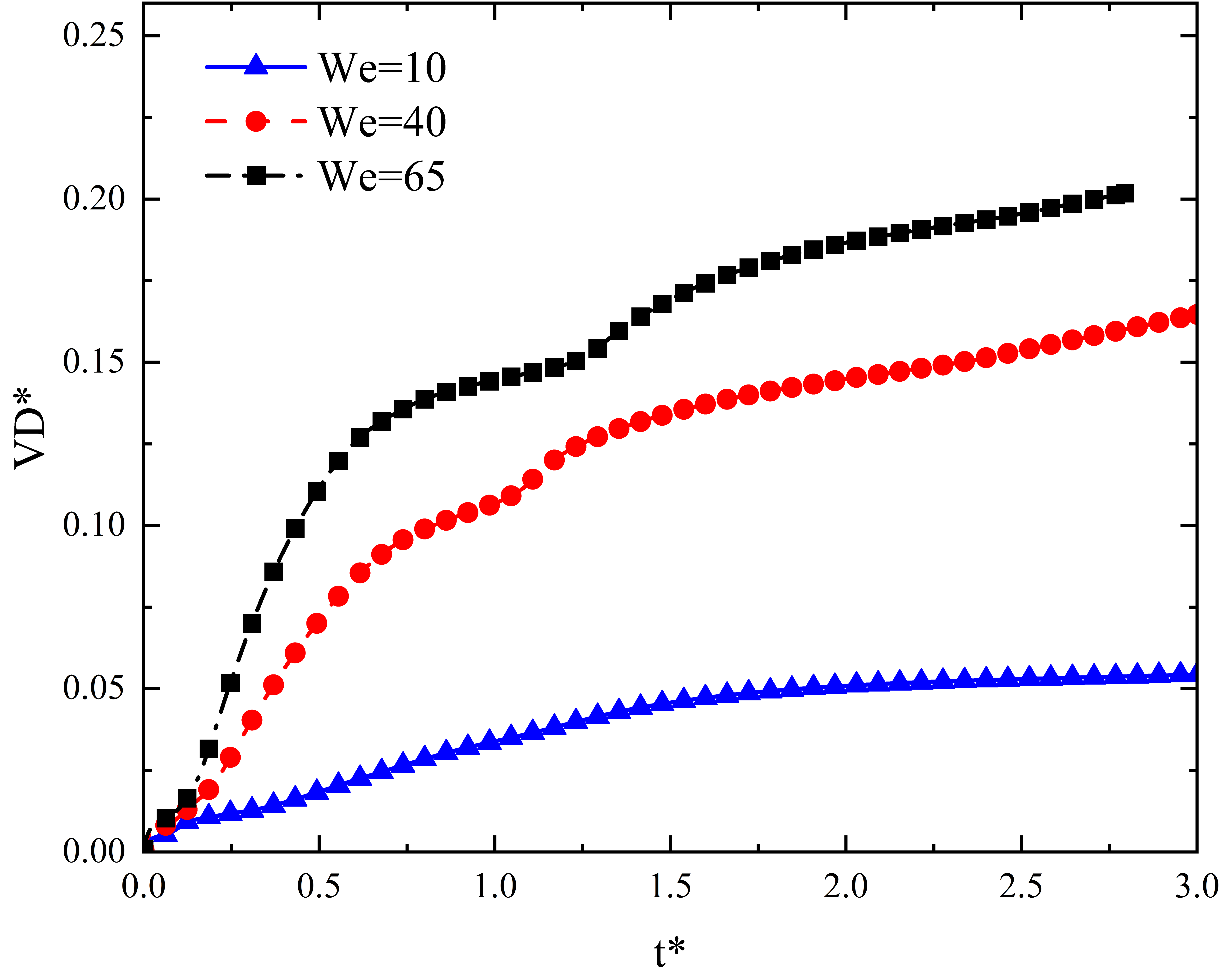}
	\caption{Temporal evolution of the dimensionless dissipated energy for different Weber numbers at $\Delta \theta  = 70^\circ$, $r/R=0.4$, and $h/R=0.5$.}
	\label{fig11}
\end{figure}

To provide deeper insight into the $We$ effect, we also calculated the normalized residual mass ${\rm{M}}_{{\rm{top}}}^{\rm{*}}$ and penetrated mass ${\rm{M}}_{{\rm{down}}}^{\rm{*}}$ during the impacting processes at various Weber numbers. As depicted in Fig. \ref{fig9a}, it's evident that, for a fixed duration, the remaining mass above the plate typically diminishes with rising Weber numbers. This is attributed to the higher kinetic energy of the impacting droplet, enabling it to swiftly pass through the orifice for higher Weber numbers, as illustrated in Fig. \ref{fig10a}. In particular, it is interesting to note that at the beginning of the rebounding phase, the rapid retracting behavior induces an increase in the residual mass, and this phenomenon is more pronounced for a higher Weber number. Referring to the images in Fig. \ref{fig9a}, one can observe that the retraction time of the droplet lengthens with an increase in the Weber number. Also, owing to the variation of the surface area, the surface energy at each Weber number is found to increase firstly and then decrease slightly during the above-mentioned retracting stage. Thereafter, due to the unbalanced Young’s force in the orifice has a tendency to push the liquid towards the region with greater wetting (bottom plate), the penetrated mass ${\rm{M}}_{{\rm{down}}}^{\rm{*}}$ increases as ${t^*} > 1.5$. Additionally, as indicated in Fig. \ref{fig10b}, under the same dimensionless time, the surface energy is increased in Weber number as a result of the increasing maximum droplet surface area. Moreover, comparing the curves in Fig. \ref{fig11}, one can conclude that the dissipated energy is enhanced with the increase of the Weber number. This is because a more significant segment of the droplet participates in the deformation for a comparatively higher Weber number, leading to this outcome. As a consequence, the fraction of the dissipated energy is also more significant. Another point that needs to be mentioned is that for a given dimensionless time, the viscous dissipation rate obtained at $We=65.0$ is always more prominent than the other two cases, especially during the spreading phase.

\subsection{Effect of pore size}
\label{section5.3}

\begin{figure}[H]
	\centering
	\includegraphics[width=0.7\textwidth]{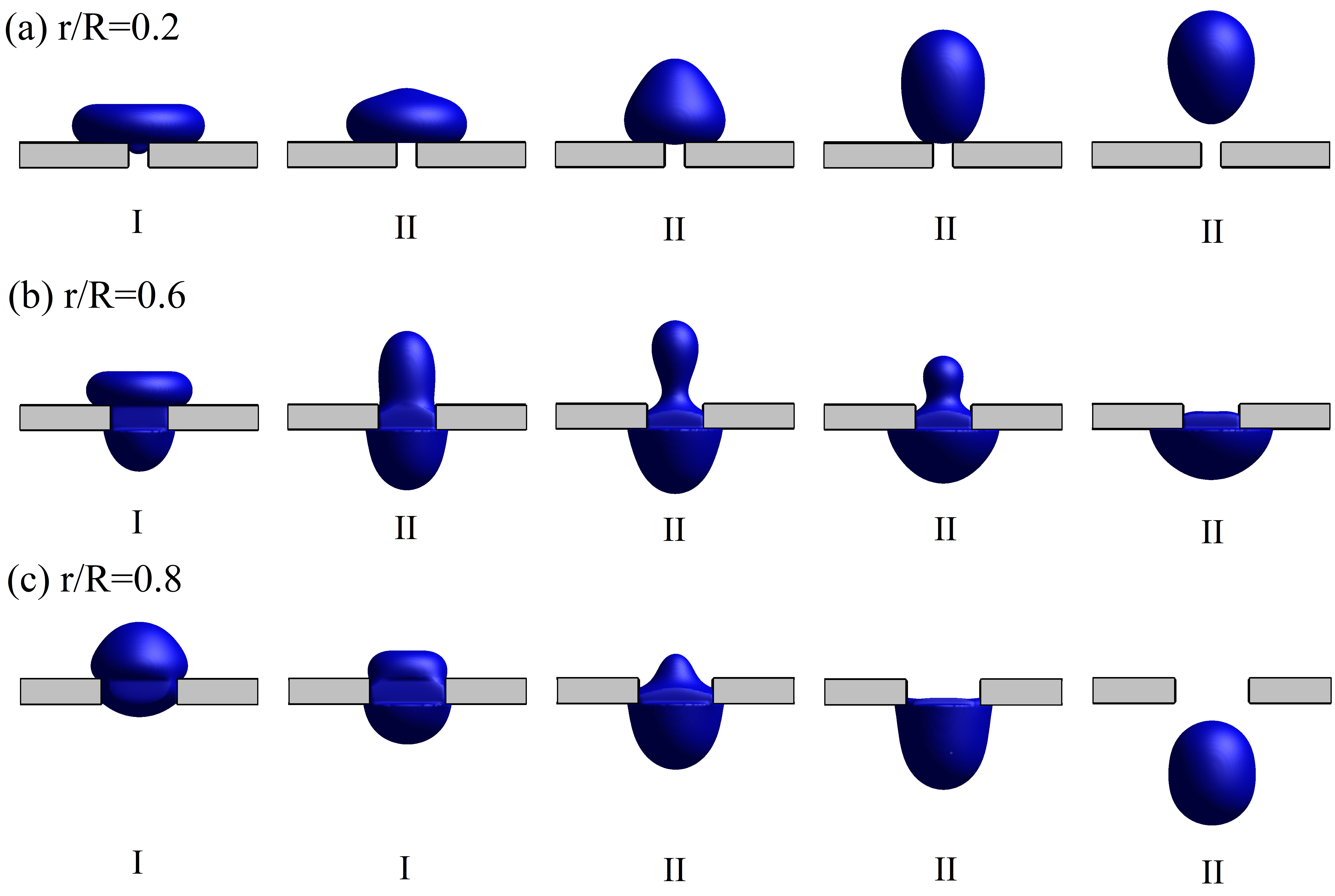}
	\caption{The droplet morphologies impacting on the perforated plate with different pore sizes $r/R$ at $\Delta \theta  = 70^\circ $, $We=10.0$, and $h/R=0.5$. I and II represent the spreading phase and rebounding phase, respectively.}
	\label{fig12}
\end{figure}

Now, attention turns to the effect of pore size $r/R$. In the simulations, the pore size $r/R$ increases from 0.2 to 0.8. The wettability difference is set to be $\Delta \theta=70^\circ$, Weber number, and the ratio between plate thickness and droplet radius are fixed at $We=10.0$, $h/R=0.5$. Fig. \ref{fig12} displays droplet morphologies on a perforated plate with varying pore sizes. It is found that the droplet firstly spreads, influenced primarily by inertial force, and subsequently rebounds above the plate due to the surface tension at $r/R=0.2$. We observe that increasing the pore radius induces droplets to penetrate the pore interior (see Fig. \ref{fig12}(b) and Fig. \ref{fig12}(c)). A fraction of the droplet passes through the pore and spreads along the bottom plate surface owing to the hydrophilic attraction (see Fig. \ref{fig12}(b)). The other part of the droplet bounces back, forming a column, which is similar to the phenomenon reported by Bordoloi et al. \cite{Bordoloi2014}. Moreover, Inertial force and the unbalanced Young's force continue to carry most of the droplet downward, and the height of the liquid column decreases constantly. As a result, the droplet attains a pendant shape in the balance of inertial force, unbalanced Young's force, and hydrophilic attraction (see Fig. \ref{fig12}(b)). Due to the relatively large pore size, the whole droplet is completely across the pore at $r/R=0.8$ (see Fig. \ref{fig12}(c)).

\begin{figure}[H]
	\centering
	\includegraphics[width=0.3\textwidth]{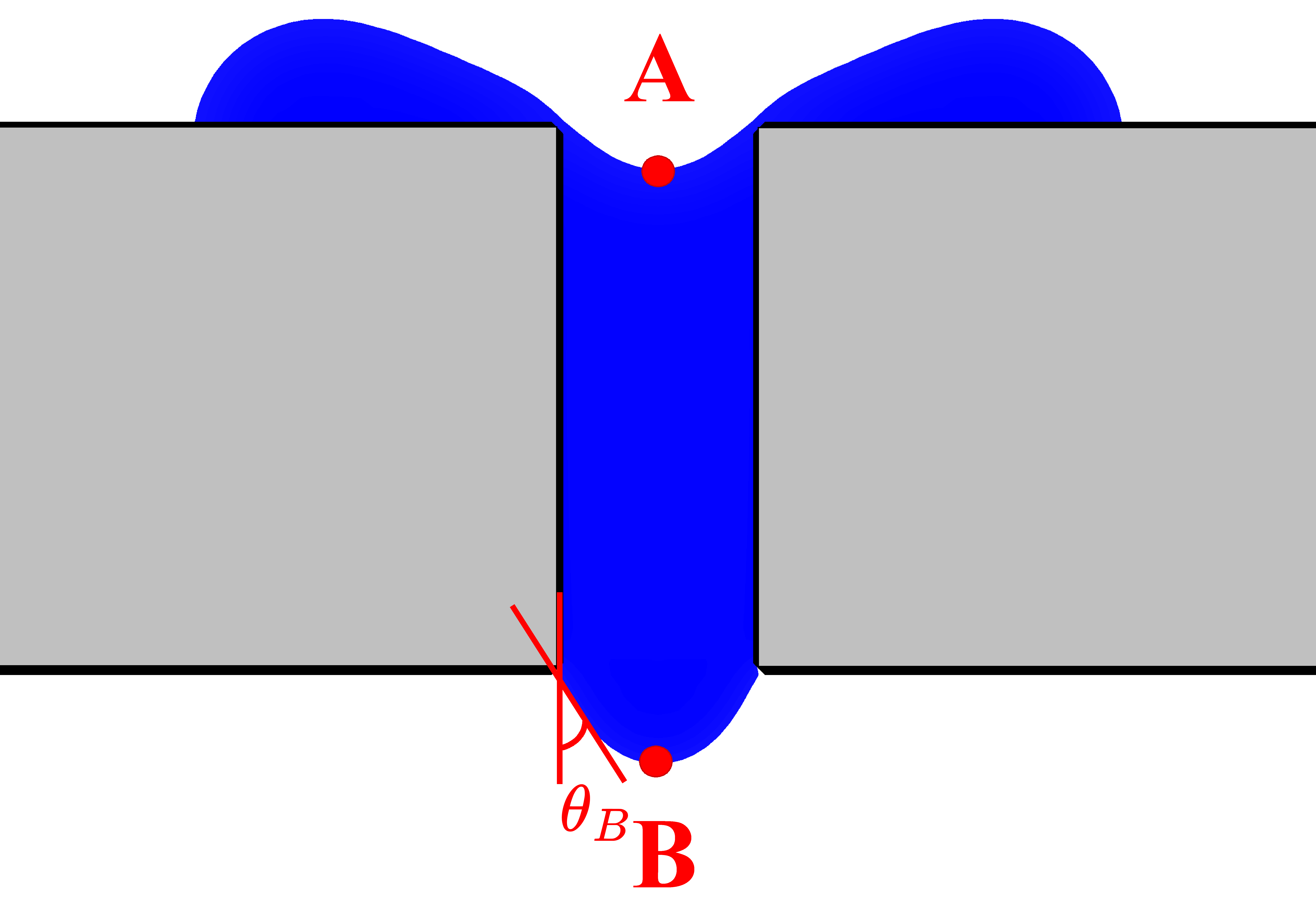}
	\caption{The ideal criticality of the dewetting phenomenon during the spreading phase.}
	\label{fig13}
\end{figure}

\begin{figure}[H]
	\centering
	\includegraphics[width=0.65\textwidth]{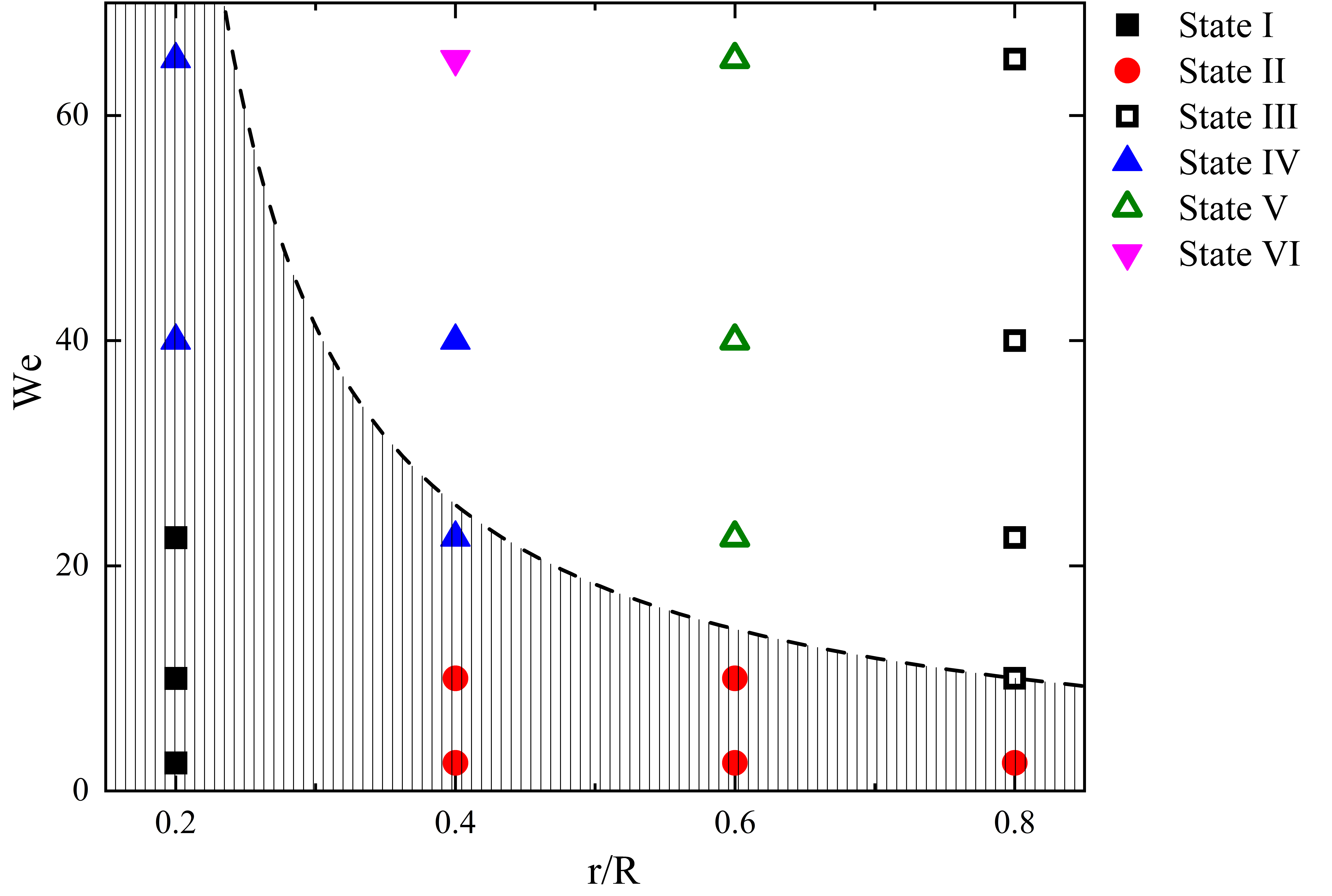}
	\caption{Phase diagram of the different states observed for droplet impacting perforated plate with a wettability-patterned orifice at $\Delta \theta  = 70^\circ $ and $h/R=0.5$. State I: droplet bouncing case, State II: droplet being captured case, State III: droplet passing through the pore case, State IV: splitting droplets bouncing and being captured case, State V: splitting droplets bouncing and passing through the pore case, State VI: splitting droplets bouncing, being captured, and passing through the pore case. The dashed line represents the critical Weber number for the dewetting phenomenon.}
	\label{fig14}
\end{figure}

\begin{figure}[H]
	\centering
	\includegraphics[width=0.65\textwidth]{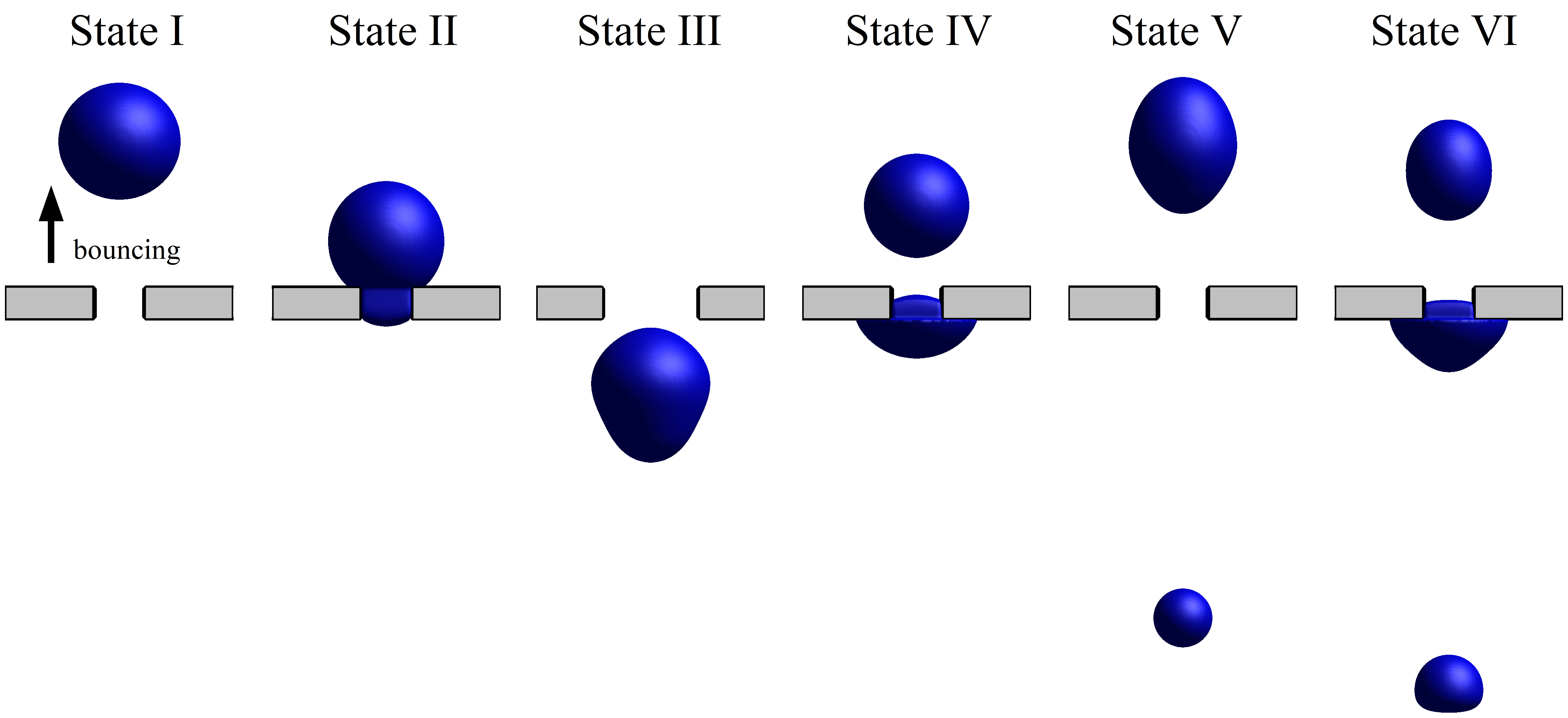}
	\caption{The droplet morphologies impacting on the perforated plate for different Weber numbers and pore sizes at $\Delta \theta  = 70^\circ$ and $h/R=0.5$.}
	\label{fig15}
\end{figure}

The above simulation results indicate that the Weber number $We$ and pore size $r/R$ play a crucial role in droplet impact behavior. Subsequently, We plotted the Weber number and pore size as variables in the phase diagram (see Fig. \ref{fig14}), in which the wettability difference and the ratio between plate thickness and droplet radius are fixed at $\Delta \theta  = 70^\circ $, $h/R=0.5$, the Weber number and pore size vary from 2.5 and 0.2 to 65 and 0.8. As shown in Fig. \ref{fig14}, there are six typical states (see Fig. \ref{fig15}) for droplets impacting at different $We$ and $r/R$. Referring to the above qualitative analysis, it can be clearly seen that the liquid finger breaks up when the dewetting phenomenon occurs. The formation of the dewetting phenomenon can be explained as a result of the competition between the inertial force, unbalanced Young's force, and the capillary. Fig. \ref{fig13} shows the critical stage of dewetting generation. At this stage, the Laplace pressure is expressed as ${{2\sigma\cos{\theta_B}}\mathord{\left/{\vphantom{{-2\sigma\cos{\theta_2}}r}}\right.\kern-\nulldelimiterspace} r}$ at the bottom interface (marked as point B in Fig. \ref{fig13}), whereas the dynamic pressure and Laplace pressure at the top of the droplet (marked as point A in Fig. \ref{fig13}) are $0.5{\rho _L}{U^2}$ and ${{2\sigma}\mathord{\left/{\vphantom {{2\sigma } r}} \right.\kern-\nulldelimiterspace} r}$, respectively. It is worth being noted that the dynamic pressure at point B is zero, and the contact angle is the static contact angle \cite{Delbos2010}. Also, according to Eq. (\ref{eq19}), the pressure losses due to viscous effect on the scale of ${{\xi h{\rho _L}{U^2}} \mathord{\left/{\vphantom {{\xi h{\rho _L}{U^2}} {\left( {2r} \right)}}} \right.\kern-\nulldelimiterspace} {\left( {2r} \right)}}$ is considered in the process \cite{Pasandideh_POF_1996,Chandra_MPS_1991}, where $\xi$ is the viscous loss coefficient. The threshold of the dewetting phenomenon can be written as

\begin{equation}\label{eq23}
	\frac{1}{2}{\rho _L}{U^2} =  - \frac{{2\sigma \cos {\theta _B}}}{r} + \frac{{2\sigma }}{r} + \frac{{\xi h}}{{2r}}{\rho _L}{U^2}.
\end{equation}
Replacing the dimensionless number $We$, $r/R$, and $h/R$ into the aforementioned equation, the critical Weber number at which the dewetting phenomenon occurs can be written as 
\begin{equation} \label{eq24}
	We = 8\left[ {\frac{{1 - \cos {\theta _B}}}{{\frac{r}{R} - \xi \frac{h}{R}}}} \right],
\end{equation}
in which the viscous loss coefficient $\xi $ and ${\theta _B}$ are chosen as 0.28 and $80^\circ$, respectively. The dashed line in Fig. \ref{fig14} represents Eq. (\ref{eq24}). Additionally, as indicated in Fig. \ref{fig14}, liquid finger break-up (Pattern V and Pattern VI in Fig. \ref{fig15}) can be observed above the dashed line.

\section{Conclusions}\label{section6}

Liquid manipulation is ubiquitous in our daily lives, and in recent times, there has been a proliferation of experimental and computational investigations aimed at exploring the dynamics of droplet impact on different surfaces. However, previous numerical simulation studies based on conditions of pore wettability homogeneity did not examine the effects of differences in pore wettability \cite{Haghani2016,Wang_PRF_2020}. This study uses a three-dimensional direct numerical simulation with the pseudopotential lattice Boltzmann method to better understand the issue and gain valuable physical insights. The numerical method is first validated against the experiment by considering a droplet impact on a perforated plate. After conducting a detailed investigation, the effects of variations in wettability difference, the Weber number, and orifice size on droplet impact dynamics are analyzed.

Observations from the current numerical findings indicate that the difference in wettability between the upper and lower surfaces of the plate results in an unbalanced vertical Young's force, impeding the upward rebound motion of the droplet. In addition, the unbalanced net force in the downward direction induces the droplet to migrate directionally to the hydrophilic area, which is different from the case of the homogeneous pore. It is interesting to note that the droplet oscillates at $\Delta \theta = 70^\circ$, and a higher wettability difference results in greater viscous dissipation energy. As for the effect of the Weber number, it is observed that a dewetting phenomenon occurs at $We=65.0$ inside the pore, and a sub-volume of the liquid finger breaks up in such a case. Further, as the Weber number increases, we note that larger deformation leads to an increase in surface energy and viscous dissipation energy. Finally, in discussing the effect of pore size, we find that a droplet may pass through a relatively large pore (i.e., $r/R=0.8$) or attain a pendant shape at the bottom of the plate due to hydrophilic attraction. Due to the significant influence of pore size and Weber number on droplet impact behavior, we also plots a phase diagram with pore radius and Weber number as variables and identifies six impact states.

\section{Appendix}

\begin{figure}[H]
	\centering
	\includegraphics[width=0.5\textwidth]{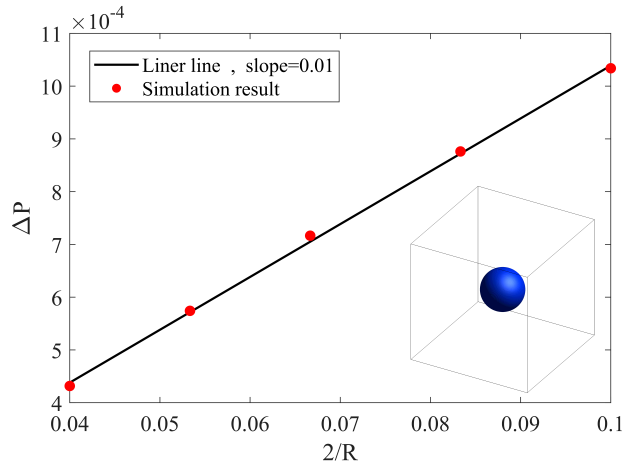}
	\caption{Verification of Laplace law}
	\label{fig16}
\end{figure}

In this appendix, two numerical tests are conducted. The first test is the so-called Young-Laplace test, which is used to show how the surface tension is determined in the pseudopotential multiphase model. According to Laplace's law, the pressure difference between the inside and outside of a droplet is linearly related to the reciprocal of the droplet's radius $R$, which in the three-dimensional case is $\Delta P = {{2\sigma } \mathord{\left/{\vphantom {{2\sigma } R}} \right.\kern-\nulldelimiterspace} R}$. In the simulation, a droplet is placed in the center of the computational domain with a grid size of $200 \times 200 \times 200$; the periodic boundary conditions are applied to all boundaries. As shown in Fig. \ref{fig16}, the LB model in this paper complies with Laplace's law, and the surface tension $\sigma$ in this work thus is fixed at 0.01.

Another test is the contact angle test, which is employed to establish the relationship between the contact angle at the solid boundary and the fluid-solid interaction coefficient $G_w$. To this end, we performed different contact angle tests by adjusting $G_{w}$ in Eq. (\ref{eq10}). In the simulation, a droplet of radius 30 is placed at the center of the bottom of the computational domain of grid size $200 \times 200 \times 200$; in the horizontal direction, a periodic boundary condition is utilized, while the top and bottom boundaries enforce a no-slip condition. Fig. \ref{fig16} shows the relationship between $G_{w}$'s different values and the equilibrium state's contact angle in the equilibrium state, and it can be seen that different contact angles can be realized by changing the solid surface $G_{w}$.

\begin{figure}[H]
	\centering
	\includegraphics[width=0.5\textwidth]{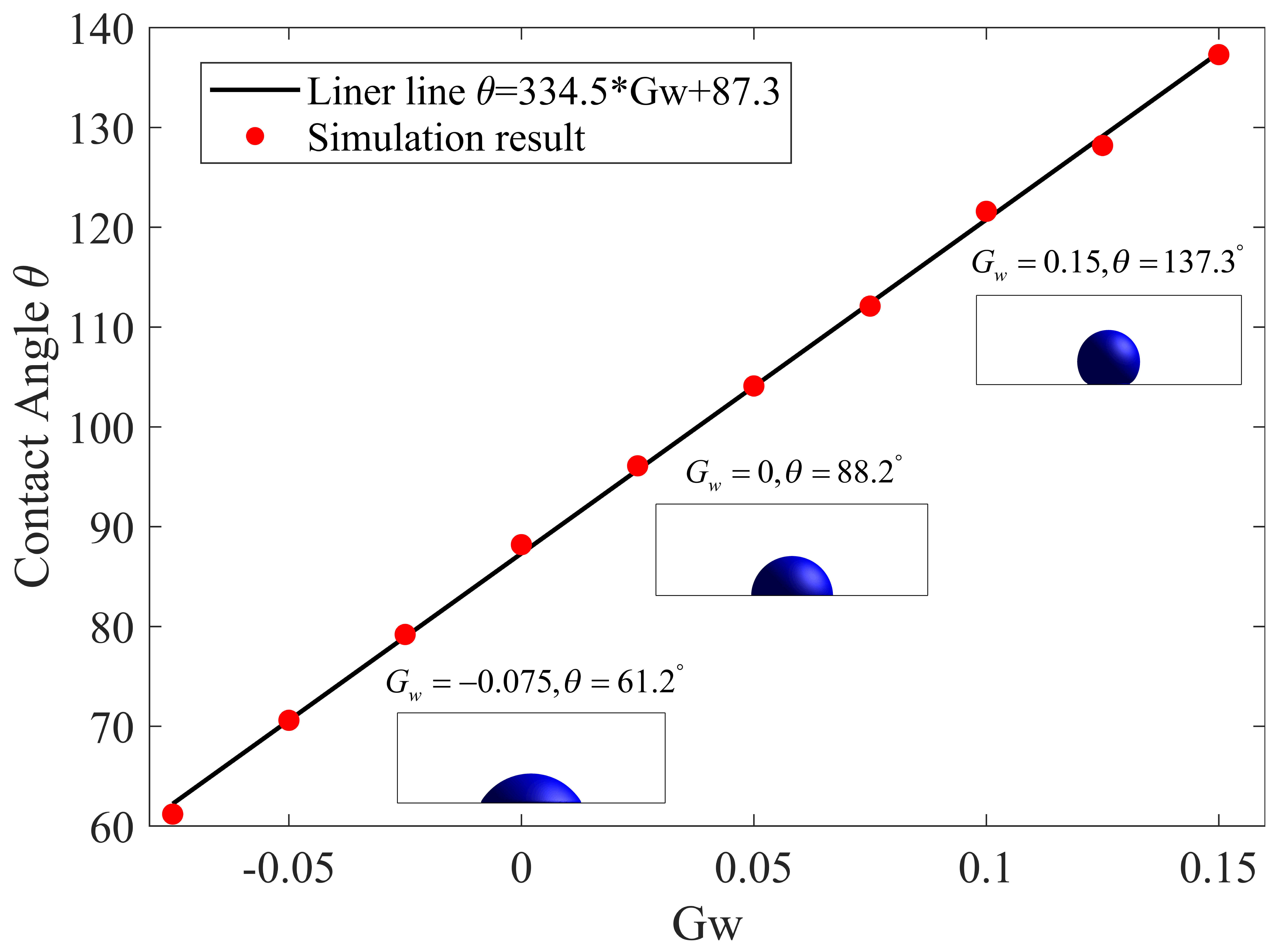}
	\caption{Contact angle $\theta$ with respect to interaction strength $G_{w}$ between the solid wall and fluid.}
	\label{fig17}
\end{figure}

\section*{Acknowledgments}

This work is financially supported by the National Natural Science Foundation of China (Grant No.  12372287).

\end{document}